\documentclass[%
reprint,
amsmath,amssymb,
aps,
pra,
superscriptaddress,
floatfix,
showkeys
]{revtex4-2}

\usepackage{color}
\usepackage{graphicx}
\usepackage{dcolumn}
\usepackage{bm}

\usepackage[mathlines]{lineno}

\usepackage{nameref}
\usepackage[colorlinks=true,     linkcolor=black,     citecolor=blue,
urlcolor=blue]{hyperref}

\usepackage[utf8]{inputenc}
\usepackage[T1]{fontenc}
\usepackage{lmodern}
\usepackage{array}
\usepackage{geometry}
\usepackage{graphicx}
\usepackage{amsmath, amssymb, amsthm}
\usepackage{bm}

\usepackage{dsfont}

\usepackage{cleveref} 
\usepackage{enumerate}
\usepackage{mathtools}
\usepackage{thmtools}
\usepackage{amsmath}
\usepackage{amsfonts}
\usepackage{thm-restate}
\usepackage{braket}
\usepackage[export]{adjustbox}
\usepackage[normalem]{ulem}

\usepackage{orcidlink}

\DeclareUnicodeCharacter{2212}{\textminus} 

\usepackage{glossaries}

\newacronym{qpu}{QPU}{Quantum Processing Unit}

\newacronym{nisq}{NISQ}{Noisy Intermediate-Scale Quantum}
\newacronym{fasq}{FASQ}{Fault-tolerant Application-Scale Quantum}

\newacronym{qec}{QEC}{Quantum Error Correction}
\newacronym{ftqc}{FTQC}{Fault-Tolerant Quantum Computation}
\newacronym{gnn}{GNN}{Graph Neural Network}
\newacronym{cnn}{CNN}{Convolutional Neural Network}
\newacronym{csmqc}{CSMQC}{Crosstalk Spectating Multiple Quantum Coherences}
\newacronym{ms}{MS}{Mølmer-Sørensen}
\newacronym{qirb}{QIRB}{Quantum Instrument Randomised Benchmarking}
\newacronym{rb}{RB}{Randomised Benchmarking}
\newacronym{mtd}{MTD}{Moving Target Defence}
\newacronym{mzlc}{MZLC}{Multi-Z-Line Control}
\newacronym{ics}{ICS}{Indirect Coupling Spectroscopy}

\newacronym{fbqc}{FBQC}{Fusion-Based Quantum Computation}
\newacronym{mzi}{MZI}{Mach–Zehnder Interferometer}
\newacronym{hom}{HOM}{Hong–Ou–Mandel}

\newacronym{gbs}{GBS}{Gaussian Boson Sampling}
\newacronym{gkp}{GKP}{Gottesman–Kitaev–Preskill}
\newacronym{rhg}{RHG}{Raussendorf Harrington Goyal}

\newacronym{aq}{AQ}{Algorithmic Qubits}

\newacronym{nv}{NV}{Nitrogen Vacancy}

\newacronym{sqc}{SQC}{Silicon Quantum Computing}

\newacronym{ccd}{CCD}{Charge-Coupled Device}

\newacronym{esr}{ESR}{Electron Spin Resonance}
\newacronym{edsr}{EDSR}{Electric Dipole Spin Resonance}
\newacronym{nmr}{NMR}{Nuclear Magnetic Resonance}

\newacronym{eom}{EOM}{Electro-Optic Modulator}
\newacronym{snspd}{SNSPD}{Superconducting Nanowire Single-Photon Detector}

\newacronym{qns}{QNS}{Quantum Noise Spectroscopy}

\newacronym{dd}{DD}{Dynamical Decoupling}

\newacronym{oqc}{OQC}{Oxford Quantum Circuits}

\newacronym{qldpc}{qLDPC}{quantum Low-Density Parity Check}

\newacronym{mot}{MOT}{Magneto-Optical Trap}

\newcommand{\etaldot}{\textit{et al.}~}

\begin{document}

\title{Crosstalk In Contemporary Quantum Devices}

\author{Spiro Gicev\,\orcidlink{0000-0003-4757-6851}}
\email{gicevs@unimelb.edu.au}
\affiliation{School of Physics, University of Melbourne, Parkville, 3010, VIC, Australia}

\author{Ben Harper\,\orcidlink{0000-0002-9192-545X}}
\affiliation{School of Physics, University of Melbourne, Parkville, 3010, VIC, Australia}

\author{Haiyue Kang\,\orcidlink{0009-0006-8222-5335}}
\affiliation{School of Physics, University of Melbourne, Parkville, 3010, VIC, Australia}

\author{Muhammad Usman\,\orcidlink{0000-0003-3476-2348}}
\affiliation{School of Physics, University of Melbourne, Parkville, 3010, VIC, Australia}
\affiliation{Quantum Systems, Data61, CSIRO, Clayton, VIC 3168, Australia}

\author{Martin Sevior\,\orcidlink{0000-0002-4824-101X}}
\affiliation{School of Physics, University of Melbourne, Parkville, 3010, VIC, Australia}

\date{\today}

\begin{abstract}
Crosstalk noise derives from phenomena in quantum devices which inhibit individual addressability or cause unintended interactions among qubits. It is widely considered one of the major problems to be solved for a quantum computing platform to operate at scales beyond one or two qubits. Despite this, detailed discussion of crosstalk is often neglected when quantum device performance is described both in the context of device benchmarking and individual algorithm execution. Additionally, while the potential for crosstalk exists in all quantum platforms, the mechanisms and severity of crosstalk between platforms varies significantly, increasing the barrier of entry associated with understanding and performing research on unfamiliar quantum platforms. While previous work focused on theoretical formalism or platform specific details, in this review article, we provide a comprehensive overview of crosstalk from quantum computing literature across a range of physical systems focusing on physical origins, methods of mitigation and known consequential security vulnerabilities. We describe multiple crosstalk mechanisms for all major quantum computing platforms, which are usually implicitly addressed through device design, tuning, and mitigation techniques. We also observe accelerating research regarding security implications, however with multiple avenues for further exploration, especially for non-superconducting systems. Together, this review provides a comprehensive entry point for researchers and industry engineers interested in understanding and addressing the challenges arising from crosstalk phenomena in modern quantum computing systems.

\end{abstract}


\maketitle


\begin{figure*}[t]
    \centering
    \includegraphics[width=1\linewidth]{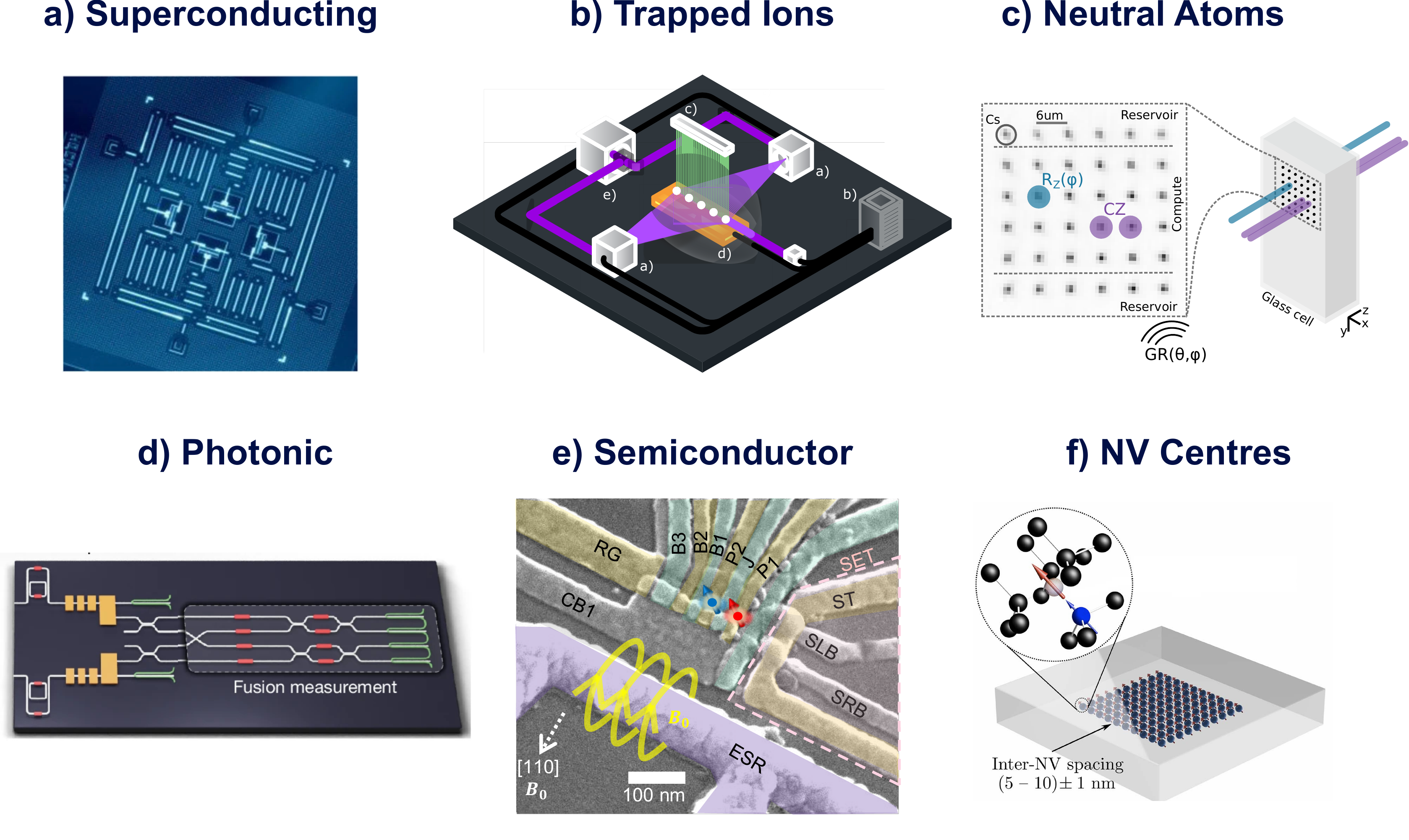}
    \caption{Quantum computing platforms. a) The superconducting platform, which includes Google, IBM, Oxford Quantum Circuits, and Rigetti, represented by an IBM superconducting circuit \cite{Gambetta_17_building}. b) The trapped ion platform, which includes IonQ and Quantinuum, represented by a schematic of a trapped ion array by IonQ \cite{Chen_24_Benchmarking}. c) The neutral atom platform, which includes Atom Computing, Infleqtion, Pasqal, Planqc, and QuEra, represented by an Infleqtion schematic \cite{Radnaev_25_universal}. d) The photonic platform, which includes PsiQuantum and Xanadu, represented by a PsiQuantum photonic circuit schematic \cite{Alexander_24_manufacturable}. e) The semiconductor platform, which includes Diraq, Intel, Silicon Quantum Computing, and QuTech, represented by a Diraq quantum dot scanning electron micrograph image \cite{Tanttu_24_assessment}. f) The nitrogen vacancy centre platform, represented by a Quantum Brilliance schematic \cite{Oberg_25_bottomup}. }
    \label{fig:platforms}
\end{figure*}

\section{\label{sec:Intro} Introduction}
Quantum computing aims to harness quantum-mechanical phenomena to achieve computational advantage in information processing tasks~\cite{Steane_98_Quantum} such as quantum chemistry~\cite{Feynman_82_simulating}, machine learning~\cite{Biamonte_17_quantum}, database search~\cite{Grover_96_fast}, and decryption~\cite{Shor_94_algorithms}. However, application-scale implementations of such algorithms remain to be experimentally verified, primarily due to qubit number and fidelity requirements commensurate with thousands of logical qubits and billions of logical operations \cite{Gidney_25_How, Caesura_26_Faster}. While quantum device noise has continuously reduced throughout the \gls{nisq}-era, it has yet to be fully established whether any platform will be able to achieve physical error rates low enough to completely avoid the additional physical qubit overheads associated with \gls{qec}~\cite{Campbell_24_series}. In either case, whether using moderate-scale \gls{nisq} devices \cite{Preskill_18_quantum} or full-scale \gls{fasq} devices \cite{Eisert_25_Mind}, pathways to quantum advantage will inevitably need to address the problems involved with simultaneously controlling large systems of physical qubits. Addressing crosstalk is a significant part of this effort, as it broadly corresponds to many of the noise and control difficulties that emerge in systems of many qubits.

In the context of \glspl{qpu}, crosstalk refers to physical mechanisms that introduce undesired correlations in errors among different qubits and quantum gates~\cite{Sarovar_20_detecting}. The term originates from classical electronics, where it refers to signal mixing between adjacent conductors. In contrast, \glspl{qpu} are designed using multiple different platforms, including superconducting circuits, trapped ions, neutral atoms, photonic systems, semiconducting platforms, and nitrogen vacancy centres, as shown in Figure \ref{fig:platforms}. Initial characterisation of \gls{qpu} noise is usually described using platform agnostic methods yielding individual qubit coherence times and gate fidelities during random operation~\cite{Knill_08_randomized, Gambetta_12_characterization}. This is usually sufficient to understand average device performance, even though individual measurement probabilities are rarely well described by independent gate noise and coherence times alone. Details of noise structure, including crosstalk, are revealed with generalised benchmarking methods~\cite{blumekohout_25_quantum} featuring increased expressivity but rapidly increasing numbers of parameters, limiting scalability.  However, as error rates are reduced and coherence times increase, structured noise effects can begin to dominate~\cite{Chen_21_exponential}. As such, experimental focus on crosstalk noise has seen recent rapid growth, in fields such as \gls{nisq} algorithm execution, fault-tolerant algorithm compilation, and secure quantum computing.

Attention to crosstalk has consistently been present in research optimising quantum processor designs \cite{Chamberland_20_topological} and publications showcasing the initial performance of newly released devices. In parallel to \gls{qpu} development, crosstalk has remained a focus among groups studying general \gls{qpu} noise \cite{Sarovar_20_detecting}, mitigation techniques \cite{Sundaresan_20_reducing}, and \gls{qec}~\cite{Fowler_14_quantifying}. Before continuing, we will briefly acknowledge prior reviews with a significant focus on crosstalk. Sarovar \etaldot \cite{Sarovar_20_detecting} gave a comprehensive overview of crosstalk noise in quantum devices, focusing on rigorous theoretical definitions. Saki \etaldot~\cite{Saki_21_survey} reviewed security concerns in quantum computers in 2021. Coupel and Farheen~\cite{Coupel_25_Security} have given a similar review in the context of cloud computing in 2025. Here, we provide an overview of crosstalk in contemporary quantum computing platforms, before presenting a survey of recent work and open problems associated with crosstalk. The existing review articles on crosstalk have primarily focused on theoretical aspects or platform specific details. Uniquely, our article covers a broad range of quantum computer architectures including superconducting, semiconductor, photonics, trapped ions, neutral atoms and NV diamonds, providing a unified source of learning physical phenomena which raises crosstalk effects in these platforms. Furthermore, we layout details about crosstalk mitigation techniques which will be helpful in future quantum devices. Another key aspect of our work is the consideration of crosstalk in a security context in which an adversary could potentially use it to disrupt or spy on quantum computing tasks in shared platforms. We provide both attack and its mitigation scenarios. 

The remainder of this text is structured as follows. We start by providing an overview of crosstalk phenomena discussed in literature for different quantum computing platforms in Sec.~\ref{sec:Platforms}, with techniques of crosstalk characterisation and mitigation discussed in Sec.~\ref{sec:Characterisation} and Sec.~\ref{sec:Mitigation}, respectively. The security vulnerabilities that arise from crosstalk phenomena are discussed in Sec.~\ref{sec:Security}. Finally, we discuss remaining open problems in Sec.~\ref{sec:Outlook} and conclude in Sec.~\ref{sec:Conclusion}.

\section{Crosstalk in Quantum Computing Platforms}\label{sec:Platforms}

Quantum processors have been constructed based on a variety of competing hardware platforms, each with variations in figures of merit such as qubit number, coherence times, gate error rates, and gate times. Crosstalk noise is no exception, with each platform possessing unique sensitivities and robustness. We encourage readers to see Appendix Section \ref{sec:Models} for a brief review of theoretical terms associated with crosstalk.

The following subsections give a brief introduction to the quantum computing platforms which have demonstrated quantum processors sufficient to perform small quantum algorithms, or are making significant progress towards that goal. We discuss the prominent crosstalk noise sources observed and expected for each platform in general, and additionally note the relevant variations present within devices of the same platform. The discussion of associated hardware-specific characterisation and mitigation strategies will also be given, with general strategies postponed to Sections \ref{sec:Characterisation} and \ref{sec:Mitigation}.

\begin{figure*}[t]
    \centering
    \includegraphics[width=1\linewidth]{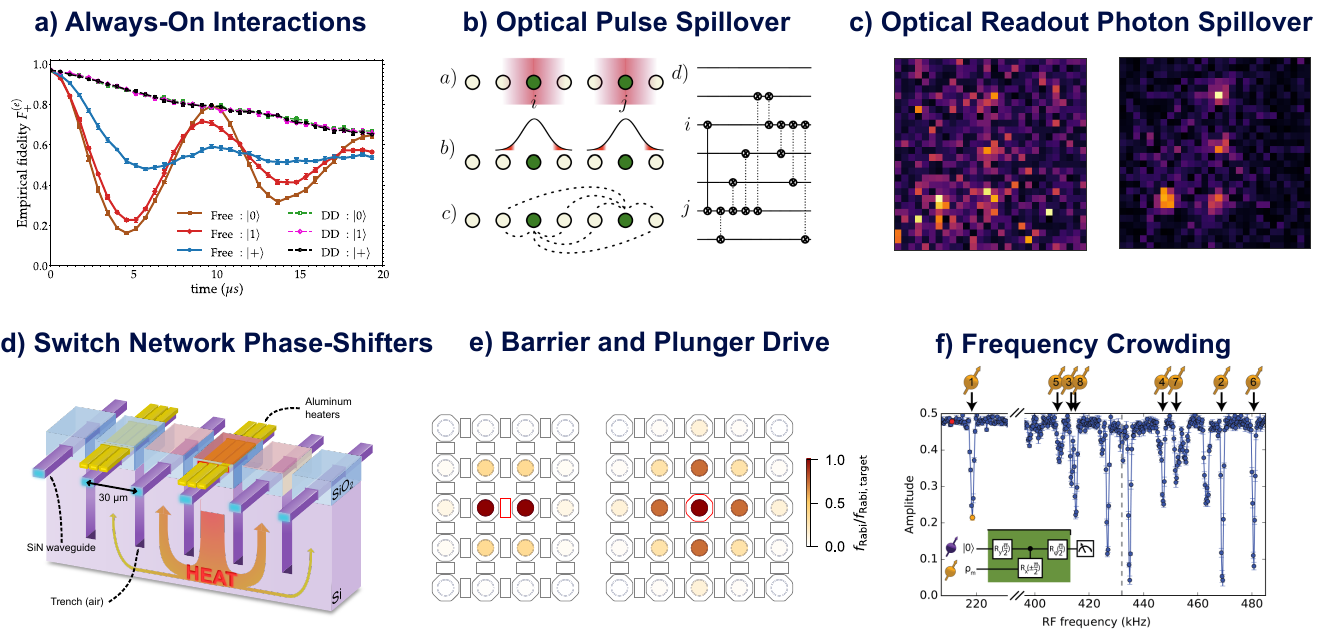}
    \caption{Examples crosstalk phenomena in \glspl{qpu} across different platforms. a) Always-on interactions causing oscillations in fidelity with dependence on neighbouring qubit states on an IBM \gls{qpu} \cite{Tripathi_22_suppression}. b) A diagram representing optical pulse spillover during \gls{ms} gates in a system of trapped ions \cite{ParradoRodriguez_21_crosstalk}. c) Two images showing a image of neutral atom readout with short (left) and long (right) exposure times \cite{Mude_25_Enabling}. d) Undesired heating of adjacent thermal phase-shifters in a photonic circuit \cite{Fyrillas_25_Resource}. e) Gate spillover for barrier and plunger drive in a QuTech quantum dot \gls{qpu} \cite{John_25_two}. f) Frequency crowding observed by nuclear spin spectroscopy in a QuTech nitrogen vacancy quantum device \cite{Bradley_19_ten}. }
    \label{fig:platform_crosstalk}
\end{figure*}

\subsection{Superconducting Devices}

Superconducting \glspl{qpu} encode qubits with superconducting materials through Josephson junctions operated at millikelvin  temperatures~\cite{Koch_07_charge, Clarke_08_superconducting, Devoret_13_superconducting, Krantz_19_quantum}. Single-qubit gates are implemented with microwave pulses \cite{McKay_17_efficient}, and two-qubit gates are implemented using couplers connecting pairs of qubits. \glspl{qpu} with fixed-frequency couplers use cross-resonance microwave pulses and always-on $ZZ$ interactions to apply two-qubit gates \cite{Rigetti_10_fully, Sheldon_16_procedure}, while two-qubit gates are directly implemented by controlling couplers in tunable coupler \glspl{qpu} \cite{Stehlik_21_tunable}. Measurement is performed using readout resonators. Providers of superconducting \glspl{qpu} include IBM, Rigetti, Google, and \gls{oqc}.

Undesired always-on $ZZ$ coupling between qubits~\cite{Magesan_20_effective} is regularly noted as a dominant contribution of crosstalk in superconducting \glspl{qpu} \cite{Murali_20_software, Niu_24_multi, Evert_24_syncopated}. This coupling is due to the energy shifts in the $\ket{11}$ state between qubits which share resonators~\cite{Ash_20_Experimental, Wei_22_hamiltonian}. In tunable coupler architectures, the $ZZ$ coupling can be somewhat controlled to be on only while the gate is being applied \cite{mckay_23_benchmarking}. On the other hand, in cross-resonance devices, the $ZZ$ coupling is always-on, and so these devices experience correlated noise even when idling. The severity of these effects depends on the relative frequencies (detuning) between qubits and couplers in the \gls{qpu}~\cite{Cai_21_impact}. Always-on interactions between the higher energy levels of superconducting qubits are also relevant \cite{Google_23_Supressing}. It is generally challenging to tune away all interactions, with interactions among neglected higher energy levels and higher order effects likely remaining. Figure \ref{fig:platform_crosstalk} a) shows neighbour-dependent coherent oscillations suggestive of always-on interactions.

Superconducting \glspl{qpu} are vulnerable to spillover crosstalk from microwave control \cite{Murali_20_software}. In Ref.~\cite{Google_25_Quantum}, microwave crosstalk was identified to be significantly non-local. Rabi ratios were not found to generally decay for pulses on further away qubits. In modelling, some error was associated with ``stray coupling crosstalk between nearest-neighbour and diagonal-neighbour qubits during parallel $CZ$ gates operation''. Spillover crosstalk during the application of two-qubit gates in superconducting platforms is regularly considered in terms of mitigation~\cite{Evert_24_syncopated, Murali_20_software} and attack vulnerabilities. Crosstalk may also manifest as undesired flux from the flux lines used to tune frequencies of qubits and couplers~\cite{Ma_25_characterizing}.

Superconducting \glspl{qpu} perform measurements by dispersively coupling to a readout resonator, which is then probed by sending a microwave pulse and inspecting the phase and amplitude of the reflected signal. This reflected signal data can then be used to classify the resonator as either originally interacted with a $\ket{0}$ or $\ket{1}$ state or higher energy states if multi-level readout is used~\cite{Gambetta_08_quantum, Magesan_15_machine}. Hothem \etaldot~\cite{Hothem_25_measuring} investigated mid-circuit measurement crosstalk on the 27-qubit superconducting \gls{qpu} ibm\_algiers using \gls{qirb}. Significant noise was found beyond what was expected based on error rates reported by simultaneous \gls{rb} characterisation. \gls{dd} using $XX$ pulses was able to show some reductions in error rates, suggesting that some discrepancy corresponded to coherent $Z$ errors. Superconducting \glspl{qpu} can also experience crosstalk associated with qubits sharing readout resonators~\cite{Gao_25_mitigating, Mude_25_efficient, Luchi_23_enhancing, Lienhard_22_deep}.

In Ref.~\cite{Chen_21_exponential}, readout crosstalk was mentioned as a significant concern, which was able to be mitigated by making readout resonators in close proximity to each other have sufficient frequency detuning. Phase corrections were applied due to expected phase shifts compared to results given by benchmarking circuits. Relatively local crosstalk was identified during repetition code circuits. The syndrome measurement data these circuits produced resulted in correlation matrix element values which are unexpected under local gate noise. Potential crosstalk sources were identified as photon crosstalk between readout resonators and flux crosstalk between qubits. Other operational crosstalk effects have also been reported in superconducting \glspl{qpu} when performing multiple quantum gates in parallel~\cite{Zhao_22_quantum, Murali_20_software}.

Overall, contemporary superconducting \glspl{qpu} have been observed to be subject to multiple different crosstalk sources. As superconducting \glspl{qpu} are relatively mature, featuring high numbers of qubits, across multiple publicly available devices, their crosstalk characteristics have been subject to substantial investigation.

\begin{table*}[t]
    \centering
\begin{tabular}{|c|l|}
\hline
Superconducting Devices & Always-on $ZZ$ interactions \cite{Magesan_20_effective, Evert_24_syncopated,Cai_21_impact, Google_23_Supressing, Murali_20_software, Ash_20_Experimental, Wei_22_hamiltonian}\\
 & Gate spillover \cite{Google_25_Quantum, Murali_20_software} \\
   & Coupler spillover \cite{Ma_25_characterizing} \\
   & Mid-circuit measurement \cite{Hothem_25_measuring} \\
   & Simultaneous gates \cite{Zhao_22_quantum, Murali_20_software} \\

\hline
Trapped Ions  & Measurement crosstalk \cite{Gaebler_21_suppression, Erickson_22_High, Feng_24_realization} \\
   & Gate spillover and aberrations \cite{Lim_25_Design, ParradoRodriguez_21_crosstalk, Kashyap_25_Crosstalk, Piltz_14_trapped} \\
   & Phonon mode crowding \cite{Landsman_19_Two} \\

\hline
Neutral Atoms & Van der Waals Interactions \cite{sharma_25_evaluation} \\
 & Gate spillover \cite{Radnaev_25_universal} \\
  & Readout pulse spillover \cite{Norcia_23_Midcircuit} \\
  & Readout detector blurring \cite{Phuttitarn_24_enhanced} \\
  & Reset shelving errors \cite{Gyger_24_Continuous} \\
 
 \hline
Photonics & Imperfect routing and delay networks \cite{Bartolucci_21_switch} \\
& Phase-shifter heating~\cite{Harris_17_Quantum, Annoni_17_Unscrambling, AghaeeRad_25_Scaling} \\
 & Phase-shifter electric fields \cite{Lomonte_21_Single} \\

\hline
Semiconducting Devices & Frequency crowding \cite{Thorvaldson_25_grovers, Edlbauer_25_11qubit, Tanttu_24_assessment, George_24_spin} \\
  & Gate spillover \cite{John_25_two, Philips_22_Universal} \\
  & Heating \cite{Undseth_23_Hotter, Unseld_25_Baseband, Xue_22_Quantum, Edlbauer_25_11qubit, Huang_24_High} \\
  & Reset crosstalk \cite{Volk_19_Loading} \\

\hline
Nitrogen Vacancy Centres & Frequency crowding \cite{Bradley_19_ten} \\

\hline
\end{tabular}
    \caption{A summary of crosstalk mechanisms in \gls{qpu} platforms mentioned in this review.}
    \label{tab:placeholder}
\end{table*}

\subsection{Trapped Ions}

Trapped-ion \glspl{qpu} encode qubits using charged atoms (ions) confined in electromagnetic traps. Due to the rapidly increasing difficulty of trapping larger ion systems, some architectures are scaling up by shuttling ions between multiple smaller arrays~\cite{Kielpinski_02_architecture}. This also allows ions to be shuttled to dedicated gate zones, separate from idling qubits. Gates are implemented with optical and microwave pulses~\cite{Olmschenk_07_Manipulation}. Non-shuttling systems instead implement individual gates using exclusively addressing beams~\cite{Hou_24_Individually}, beam-steering, and multiplexing. Measurement is usually performed by applying laser light to the ion to be measured, which, conditioned on the state of the qubit, may excite the atom to a higher energy level. Measurements of emitted photons from the ion are then used to identify the original state. The two-qubit gate of trapped-ion systems is the \gls{ms} gate \cite{Sorensen_99_Quantum}, a phonon-mediated interaction known for potential all-to-all connectivity. Providers of trapped ion \glspl{qpu} include IonQ~\cite{zhao_25_Quantum, Chen_24_Benchmarking, Wright_19_Benchmarking} and Quantinuum~\cite{Pino_21_demonstration, Moses_23_racetrack}, which have focused single ion chains and shuttling architectures, respectively.

Stray photons emitted during qubit measurement are a dominant source of crosstalk in trapped-ion devices. This is particularly relevant in the context of communication protocols, where the entanglement between trapped ions and emitted photons is treated as a resource. Multiple techniques have been developed to address undesired reabsorption of photons in trapped ions, such as hiding ions from measurement light~\cite{Gaebler_21_suppression} or performing measurements at dedicated gate zones. Another strategy involves transferring qubit information to another species of ion which can be subsequently measured using light which is off resonance with the first ion~\cite{Erickson_22_High}. In Ref.~\cite{Feng_24_realization} unique frequencies for two types of qubits is achieved for trapped ions of a single species by using qubits defined in $S$ and $F$ hyperfine structure levels. In systems with multiple readout ions, scattered light during readout may still cause measurement errors if all readout ions are of the same species, but would still have the benefit of not destroying information stored in the data ion species. 

General pulse spillover and aberration problems also exist for gates in trapped-ion \glspl{qpu}~\cite{Lim_25_Design}. Spillover can be understood as a partial application of a gate to neighbouring ions \cite{ParradoRodriguez_21_crosstalk}, as shown in Figure \ref{fig:platform_crosstalk} b). Next-next-neighbour spectator qubits see exponentially suppressed crosstalk errors, due to the sufficiently narrow Gaussian profile of the concentrated light pulses. Spillover can be minimised by spacing ions further apart, which also reduces the strength of two-qubit interactions, increasing gate times and demands on coherence times. Alternatively, spillover can be reduced by using narrower beams, which may cause polarisation to be ill defined and introduces errors from relative ion-beam motion~\cite{Kashyap_25_Crosstalk}.  In Ref.~\cite{Piltz_14_trapped} authors propose a way to interact trapped ion qubits using unique frequencies for each qubit. However, such techniques are at the expense of mechanical fabrication cost, and difficulty in experimental implementation. 

Finally, trapped ion \glspl{qpu} encounter phonon frequency mode crowding problems associated with their \gls{ms} gate. This causes increased gate errors when interactions are performed between multiple adjacent pairs of qubits \cite{Landsman_19_Two}.

\subsection{Neutral Atoms}

Neutral atom \glspl{qpu} encode qubits in arrays of uncharged atoms held in place using laser or magnetic fields~\cite{Wintersperger_23_neutral, Wurtz_23_aquila}. Computational basis states usually correspond to hyperfine electronic ground states. Atoms are initially trapped in \glspl{mot} and are cooled using Doppler cooling. Atoms are subsequently trapped more precisely in optical traps or moved with mobile traps using acousto-optical deflectors. Atom positions can be verified by applying appropriate frequency light and capturing an image using a \gls{ccd} camera. Single qubit gates can be applied by applying either microwave pulses resonant with the qubit transition, or by instead driving Raman transitions. Multi-qubit interactions are facilitated by exciting atoms to highly energetic (Rydberg) states~\cite{Urban_09_observation}. Developers of neutral atom \glspl{qpu} include QuEra, Atom Computing, Pasqal, Infleqtion, and Planq.

Due to the small interatomic spacing on the order of $\mu\mathrm{m}$, control pulses have a spatial extent such that they can simultaneously drive transitions in multiple adjacent atoms. This can be mitigated by shuttling qubit away to interaction zones~\cite{Muniz_24_High, Muniz_25_Repeated, Reichardt_25_Fault} or by applying addressing laser beams to shift qubit transition frequencies into or out of resonance~\cite{Radnaev_25_universal}. The focus sizes of the addressing beams are usually small enough to eliminate crosstalk which may occur in this process, though smaller separations are desirable in order to store more qubits, have stronger multi-qubit interactions and alleviate the demands on beam steering.

Measurement of neutral atoms is usually performed by stimulating atoms with resonant light and forming a picture using a \gls{ccd} camera. Qubit states need to then be classified as either bright or dark states. Usually, light emitted by bright atoms results in increased pixel intensity values near the position of the atom. However, in some cases, light can also result in increased intensity in pixels corresponding to other atoms, as shown in Figure \ref{fig:platform_crosstalk} c). This can result in erroneous classification in both the bright atom and the dark atom and corresponds to measurement crosstalk. This effect is more pronounced for short readout gates utilising short, but high intensity, readout pulses \cite{Phuttitarn_24_enhanced}. Additionally, care must be taken that undesired measurements are not performed. Atom Computing implemented techniques for addressing such measurement crosstalk by using ``hiding light'' to hide qubits which are not intended to be measured from imaging light~\cite{Norcia_23_Midcircuit}.

Planqc have addressed the problem of atom reloading in neutral atom systems by using low-loss fluorescence imaging and recycling atoms for use in subsequent shots~\cite{Gyger_24_Continuous}. While new atoms are loaded, atoms which are planned to be retained are ``shelved'' by being transitioned to a meta-stable state insensitive to the \gls{mot} used for atom loading. This constitutes a potential source of operational crosstalk for algorithms which require mid-circuit reloading.

Finally, in neutral atom devices acting as quantum simulators, Van der Waals interactions can be considered as uncontrolled always-on interactions \cite{sharma_25_evaluation}.

\subsection{Photonics}

Photonic \glspl{qpu} encode quantum information in degrees of freedom of light~\cite{Kok_07_Linear, Braunstein_05_Quantum}. Operations in photonic devices consist of optical elements including beam-splitters, phase-shifters, and continuous variable operations such as displacements, phase-space rotations, and squeezing. Photonic systems are known for difficulties involving gate implementation, which sometimes are non-deterministic and require post-selection. Errors are biased towards erasure (photon loss). Two well-known developers of photonic \glspl{qpu} are PsiQuantum and Xanadu, which develop discrete-variable and continuous-variable designs respectively.

PsiQuantum are developing fault-tolerant \glspl{qpu} based on \gls{fbqc} in dual-rail photonic systems~\cite{Bartolucci_23_fusion}. Computational basis states correspond to a photon present on one of the two rails. \gls{fbqc} uses resource state generators which route output qubits to fusion devices, where destructive joint measurements are performed on pairs of qubits. The interaction between generators in space and time form fusion networks capable of implementing \gls{qec} codes. 

A potential source of crosstalk within \gls{fbqc} devices corresponds to switching faults within switch networks~\cite{Bartolucci_21_switch, Mendoza_15_active}, which mediate  multiplexing, de-multiplexing, and delays. The fundamental building block of switch networks is the \gls{mzi}~\cite{Lahiri_16_basic}, which implements swaps on two inputs depending on the value given to a phase-shifter. This phase depends on physical properties, such as phase-shifter dimensions, temperature or voltage. PsiQuantum have implemented switch networks using thermal phase-shifters and demonstrated high fidelity single qubit and two-qubit operations. Thermal phase-shifters have been identified as a source of crosstalk in prior programmable nanophotonic circuits~\cite{Harris_17_Quantum, Annoni_17_Unscrambling, Fyrillas_25_Resource}, as illustrated in Figure \ref{fig:platform_crosstalk} d). PsiQuantum's next generation devices are planned to use barium titanate (BTO) phase-shifters, based on the Pockels effect~\cite{Hecht_02_optics}, for higher-speed switching~\cite{Alexander_24_manufacturable}.

Although experimental literature currently lacks  results involving BTO phase-shifter switch networks, results do exist for other Pockels phase-shifters. 
Lomonte \etaldot~\cite{Lomonte_21_Single} provided results of photon switching performance of single photon \glspl{mzi} implemented in Lithium-Niobate-On-Insulator (LNOI) using a lithium niobate \glspl{eom}. The system included one \gls{eom} and two \glspl{snspd}. During DC operation, a voltage of 17.8V was found to achieve a $\pi$ phase shift, completely switching all light from one detector to the other, with a count rate stable across 12 hours of operation. In high speed AC operation, electrical crosstalk between the \gls{eom} and \glspl{snspd} was identified to limit adequate operation to below 1 GHz, with crosstalk induced oscillations in detector readout at 1 MHz. This was attributed to the dual-purpose RF probe employed in measurements and the close proximity of \gls{eom} and \glspl{snspd} contact pads, and was mitigated with frequency filters. Increasing the spatial separation of the \gls{eom} and \gls{snspd} and separating the \gls{eom} and \glspl{snspd} driving channels onto different RF
probes were suggested as potential improvements.

Xanadu is building fault-tolerant \glspl{qpu} based on hybrid resource states~\cite{Bourassa_21_blueprint}. In particular, \gls{gbs} devices act as resource state generators which produce heralded \gls{gkp} states~\cite{Gottesman_01_Encoding} of light with low probability, but can be boosted by multiplexing~\cite{Tzitrin_20_progress}. \gls{gkp} states are known to have been produced when the output of the \gls{gbs} devices produces a particular measurement pattern in the associated photon number resolving detectors, which are the only system components with non-room temperature designs so far. Momentum squeezed vacuum light states are used when \gls{gkp} state preparation fails. Quantum information is encoded exclusively in the \gls{gkp} states, while encoded information is processed using measurement-based quantum computing on hybrid \gls{gkp} and squeezed states. Homodyne measurements are able to be applied to these states, which are argued to be able to be some of the fastest measurements compared to other photonic and non-photonic platforms. Optical delays can be used to prepare repeated $\ket{+}$ GKP states output by high repetition rate \gls{gbs} devices into 1D cluster states by applying $CZ$ gates on qubits sequentially. When multiple \gls{gbs} devices are arranged in 2D, and $CZ$ gates are available with nearest-neighbour connectivity, 3D cluster states such as those defined by the \gls{rhg} lattice may be constructed. In both the 2D case and 3D case, the final dimension is time, and only subsequent pairs of qubits from each \gls{gbs} device are required to exist in any instant of time.

Xanadu have shown experimental distance-2 repetition code performance using cluster states~\cite{AghaeeRad_25_Scaling}. \gls{gbs} chips were implemented with thermo-optic phase shifters. Thermal crosstalk was present and was mitigated by iterative calibration.

\subsection{Semiconducting Devices}
Semiconductor \glspl{qpu} build upon classical semiconductor technological developments, aiming for eventual large-scale fabrication~\cite{Schofield_25_roadmap}. Semiconductor \glspl{qpu} generally fall under two categories corresponding to the Kane proposal~\cite{Kane_98_Silicon}, which encodes information in dopant atoms, and the Loss and DiVincenzo proposal~\cite{Loss_98_Quantum}, which encodes information in electrons trapped in electric fields (quantum dots). Qubits are usually controlled using a combination of magnetic fields for Zeeman splitting and oscillating magnetic or electric fields for manipulating the spin via \gls{esr} or \gls{edsr}. Developers of semiconductor \glspl{qpu} include \gls{sqc} (phosphorous dopants), and Diraq, Intel, and Qutech (quantum dots).

A potential source of operational crosstalk in semiconducting \glspl{qpu} corresponds to frequency crowding. For example, in \gls{sqc} devices, single qubit operations are performed with a broadband antenna implementing \gls{esr} and \gls{nmr} with radio frequency and microwave signals respectively~\cite{Thorvaldson_25_grovers, Edlbauer_25_11qubit}. The nuclei-dependent electron spin resonance frequency results in native gates corresponding to multiply-controlled-$Z$ operations, controlled on each nucleus in each cluster. This introduces the potential for crosstalk/correlated errors in order to perform operations controlled on fewer spins. A single antenna was also used when investigating the performance of a device with two adjacent nuclear spin registers. This resulted in increased demands for frequency separation/resolution (48 resonances).

Quantum dot \glspl{qpu} are also vulnerable to frequency crowding. Diraq constructs quantum devices using electron spin quantum dots in silicon. Diraq investigated the noise present in 3-dot and 4-dot chains by applying gate set tomography methods~\cite{Tanttu_24_assessment}. It was concluded that, although qubit frequency separations were large compared to the Rabi frequencies, driving errors were caused by crosstalk due to insufficient frequency separation. Similar devices by Intel used a micromagnet, in conjunction with an externally applied magnetic field, to separate the resonant frequencies of the qubits, allowing individual control of qubits using a shared microwave source \cite{George_24_spin}.

QuTech have investigated local driving crosstalk for a device containing an array of 10 quantum dots~\cite{John_25_two}, with projections given for larger systems, as shown in Figure \ref{fig:platform_crosstalk} e). Detailed results were shown for different hole occupations taking values of 1, 3, and 5, which were shown to significantly change the amount of operational crosstalk experienced. It was also observed that barrier drive crosstalk was lower for hole occupations of 1, while hole occupations of 3 caused lower plunger drive crosstalk. Overall, crosstalk Rabi ratios were observed on the order of 10\%. This is consistent with previous results observed in smaller devices~\cite{Philips_22_Universal}. These effects are consistent with correlated fluctuations in electric fields at qubit positions. 

QuTech documented a temperature dependence of silicon spin-qubit frequencies and note that it weakens at increasing temperature~\cite{Undseth_23_Hotter}. The precise physical mechanism was not concluded by this work. It is argued that this temperature dependence in frequency shifts is consistent with previous heating effects associated with microwave control. Low frequency, (Baseband) control was suggested as a means of reducing crosstalk associated with microwave fields, with gates mediated by moving electrons between dots~\cite{Unseld_25_Baseband}. In previous work, the heating-based crosstalk effect was mitigated by applying compensating $Z$ rotation phase corrections~\cite{Xue_22_Quantum}. In Ref.~\cite{Edlbauer_25_11qubit} \gls{sqc} observed a dependence of electron spin and nuclear spin frequencies on microwave drive amplitude was observed. This effect was mitigated by ensuring consistent input power during microwave antenna operation during partial or complete idling periods. Diraq also investigated the errors present in a two-qubit quantum dot device as a function of system temperature~\cite{Huang_24_High}. Extensive detail was given, including discussions of crosstalk. System errors associated with a $CZ$ gate were analysed with pyGSTi, showing both Hamiltonian (coherent) and stochastic (incoherent) components.

QuTech have previously observed that loading large quantum dot systems experience significant capacitive crosstalk. A tuning procedure was given to mitigate crosstalk and is argued to be scalable~\cite{Volk_19_Loading}. 

\subsection{Nitrogen Vacancy Centres}
\gls{nv} centres are optically active point defects in diamond corresponding to a substitutional nitrogen atom with an adjacent lattice vacancy~\cite{Doherty_13_nitrogen}. \glspl{nv} are distinguished amongst other quantum technologies by their ability to remain coherent in ambient laboratory conditions. Qubits can be encoded in the ground state electronic structure of \glspl{nv} and the nuclear spin states of the substitutional nitrogen or adjacent $^{13}$C nuclei \cite{Waldherr_14_quantum, Bradley_19_ten}. Electron and nuclear spins can be manipulated with microwave and radio-frequency pulses respectively. \gls{nv} electron readout and initialisation is achieved by pumping with green laser light, while nuclear spin readout and initialisation is achieved indirectly by manipulating the electronic state of the \gls{nv}. Multi-qubit gates for nuclear spin qubits within range of an \gls{nv} can be mediated by microwave transitions of the electron of the \gls{nv}, providing efficient multi-qubit controlled phase gates within each cluster of nuclei. Multiple techniques have been proposed for mediated interactions between adjacent \glspl{nv}~\cite{Oberg_24_atom, Dolde_13_room, Yao_12_scalable, Doherty_16_towards}. Quantum Brilliance has developed \gls{nv} \glspl{qpu} which implemented small scale \gls{nisq} algorithms within standard modern supercomputing racks~\cite{herrmann_23_quantum, Nguyen_24_software}.

Frequency crowding is a major source of crosstalk in \gls{nv} \glspl{qpu}. Bradley \etaldot~\cite{Bradley_19_ten} observed frequency crowding crosstalk when investigating a ten-qubit NV quantum register, consisting of an electron spin, one $^{14}\mathrm{N}$ nuclear spin of an NV and eight $^{13}\mathrm{C}$ nuclear spins in the diamond lattice. The nine nuclear spin resonances were observed over a range between 220 kHz and 480 kHz, with non-uniform spacing and some overlap, as shown in Figure \ref{fig:platform_crosstalk} f). Device performance was benchmarked by preparing entangled states. When investigating the fidelity of GHZ sets of increasing numbers of qubits, crosstalk was identified as the cause of a significant fidelity reduction beyond what would be expected from a independent depolarising noise model. Noise models which included crosstalk effects yielded mixed success at matching Bell state fidelities observed experimentally. Frequency resolution is addressed in Quantum Brilliance's strategy for realising large-scale arrays of NV qubits in diamond via a bottom-up fabrication process~\cite{Oberg_25_bottomup}. A gradient magnetic field is proposed to differentiate the electron spin resonances for different NV centres. This results in a potential source of crosstalk deriving from disorder in the spatial positions of NV centres, which would place electron spin resonances closer than intended. It has been suggested that placement accuracy of $\pm1$ nm is sufficient to provide sufficient spectral differentiation~\cite{chen_24_optimisation}.

\section{Crosstalk Characterisation and Detection}\label{sec:Characterisation}
Characterisation of crosstalk corresponds to either fitting a noise model which includes crosstalk errors (see Section \ref{sec:Models}) or identifying particular properties of any crosstalk that is present. This requires explicit definitions of protocols which allow conclusions to be made about the noise of a \gls{qpu}.

\subsection{Component Characterisation}
A standard method of characterising crosstalk is to perform simultaneous \gls{rb} and compare with independent runs of \gls{rb} \cite{Gambetta_12_characterization, Arute_19_quantum}. Any observed increase in error rates can be directly attributed to crosstalk. Care must be taken to ensure that drift is not being confused with crosstalk. Hothem \etaldot~\cite{Hothem_25_measuring} designed and implemented a modified \gls{rb} protocol, \gls{qirb}, which is sensitive to crosstalk noise in the Quantinuum H1-1 device. The crosstalk corresponds to stray light from measurement beams or fluorescing ions which can interact with unmeasured ions. Long range crosstalk was found to exist across different gate zones in addition to within gate zones. Xia~\etaldot~\cite{Xia_15_randomized} characterised crosstalk while \gls{rb} circuits were performed on a qubit and found average crosstalk error rates of $0.002\pm0.009$ across an entire 47 site array, $0.014\pm0.02$ for nearest-neighbour qubits and $0.0005\pm0.001$ for non-nearest-neighbour qubits. In comparison, when simultaneous identical \gls{rb} was performed, which is insensitive to the spillover problem for identical atoms, an error rate of $0.0018\pm0.0014$ was achieved, suggesting that nearest-neighbour crosstalk was significant for this system, but non-nearest-neighbour crosstalk was not more prominent than the noise sources present in simultaneous identical \gls{rb}.

Directly performing process tomography provides some details about crosstalk present in a \gls{qpu}. It scales badly with system size. Idle tomography \cite{BlumeKohout_19_idle} is a special case, where it not only has the freedom of tuning the level of locality and channels to characterise, but also being highly compact that data from one tomography experiment can be shared by multiple analyses of channel parameters. In fact, the number of quantum experiments only scales in $O(\log(n))$ for $n$ qubits, yet classical post-processing still scales exponentially to obtain the desired parameters. As the name suggests, Idle Tomography is initially designed to characterise noise parameters of a noisy idling quantum channel. To characterise gate-induced crosstalk noise, one needs to compare the channel parameters sampled from purely idling circuits against idling circuits with nearby crosstalk gates. Idle tomography assumes the noise is small (so the corresponding process matrix $\Lambda\coloneq e^{\mathcal{L}}\approx \mathds{1}+\mathcal{L}+O(\mathcal{L}^2)$ can be approximated to first order Taylor expansion of the Lindbladian generator), and Markovian \cite{BlumeKohout_22_Taxonomy}. In addition, it assumes the noise comprises multiple channels, including coherent rotations, Pauli error, non-unital channels, etc., which are characterised separately. Idle tomography also gives the freedom to choose the largest $k$-local representation of each channel to be characterised, e.g., only single and two-qubit unitary generators are characterised for coherent channels for $k=2$. Here, we give a brief demonstration of how idle tomography is performed for the Pauli channel on one qubit. The first step is to prepare the qubit in one of the Pauli eigenstates, e.g., $\ket{0}$ as the +1 Pauli-$Z$ eigenstate. Then the qubit is left idled for $L$ clock cycles while the crosstalking gates are acted on the nearby qubits for $L$ times. Under Pauli channels, $\ket{0}$ is only affected by the Pauli-$X$ and Pauli-$Y$ channels with observed probability of $r_Z=\epsilon_X+\epsilon_Y$ of flipping it into $\ket{1}$, where $\epsilon_X$, $\epsilon_Y$ are the Pauli-$X$ and Pauli-$Y$ channel error rates. In practice, the probability $r_Z$ can be obtained from the gradients of the linear extrapolation of the proportion of flipped qubits versus $L$. After trailing this for all Pauli eigenstates, one can solve the equation
\begin{equation}
    \begin{bmatrix}
                r_X \\
                r_Y \\
                r_Z
        \end{bmatrix} =\begin{bmatrix}
            0 & 1& 1\\1 & 0 & 1\\ 1 & 1 & 0
        \end{bmatrix}\begin{bmatrix}
            \epsilon_X\\
            \epsilon_Y\\
            \epsilon_Z
        \end{bmatrix}
\end{equation}
for $\epsilon$ to obtain the corresponding Pauli channel parameters. It is remarked that the procedure of characterising Pauli channels above always prepares and measures the qubit in the same basis, which are sensitive to stochastic errors but not coherent errors to the first order of the Lindbladian. This explains why noise parameters for different channels can be sampled separately without interference with each other.

Crosstalk can be present when multiple measurements are performed in parallel. Quantum detector tomography can be used to verify whether individual measurements are indeed being performed as expected~\cite{Seo_22_measurement}.

\subsection{Detection}
Crosstalk detection can be a viable alternative when complete characterisation is impractical. It corresponds to preparing circuits or states which give measurement results which confirm the existence of crosstalk events with high certainty.

Rudinger \etaldot \cite{Rudinger_19_probing} developed a general protocol for determining the presence of context dependence given the result of circuit measurements. Although crosstalk corresponds to one form of context dependence, other phenomena such as drift may also be present. Efforts can be made to rule out drift to allow conclusive statements to be made regarding crosstalk in particular.

Kang \etaldot~\cite{Kang_25_time} investigated the capability of detecting crosstalk attacks with multiple quantum coherences and developed the protocol of crosstalk-spectating multiple quantum coherences (CSMQC) to post-select shots where no attack was detected. Hence, this is a form of crosstalk detection which can be performed \textit{in situ}, detecting crosstalk events soon after they occur.

\subsection{Figures of Merit}
Simplified descriptions of crosstalk-related errors can be used in cases when complete descriptions of noise channels are impractical. For example, a in terms of control pulses, the Rabi ratio can be an efficient manner to describe the extent of crosstalk present. Calculating Rabi ratios corresponds to measuring the ratio of the Rabi frequency of neighbouring qubits compared to pulsed qubits. 

The Infleqtion trapped-ion device discussed in Ref.~\cite{Radnaev_25_universal} showed signs of crosstalk (possibly pulse spillover), with contributions both from the expected pulse spillover and anomalous effects. A value of $0.3\%$ was observed for the ratio of pulse intensity at neighbouring atoms with respect to target atoms. Assuming a Gaussian beam shape, a value 10 times smaller was expected, suggesting additional unknown effects are present.

\subsection{Bounds}
In some cases, limited results are available regarding crosstalk in particular components of some quantum devices. In these cases bounds on crosstalk can sometimes still be calculated. For example, while the potential for crosstalk during the use of \glspl{mzi} switches has seen limited study, one figure of merit characterising the performance of \glspl{mzi} is the \gls{hom} quantum interference, which has been measured for PsiQuantum devices using thermal phase-shifters~\cite{Alexander_24_manufacturable}. It should be noted that \gls{hom} quantum interference visibility is also sensitive to indistinguishability, spectral purity, number purity, signal-to-noise ratio, and system detection efficiency. Therefore, this measurement alone can only provide upper bounds on individual non-idealities such as crossstalk. 

\section{\label{sec:Mitigation} Crosstalk Mitigation}

\begin{figure*}
    \centering
    \includegraphics[width=1\linewidth]{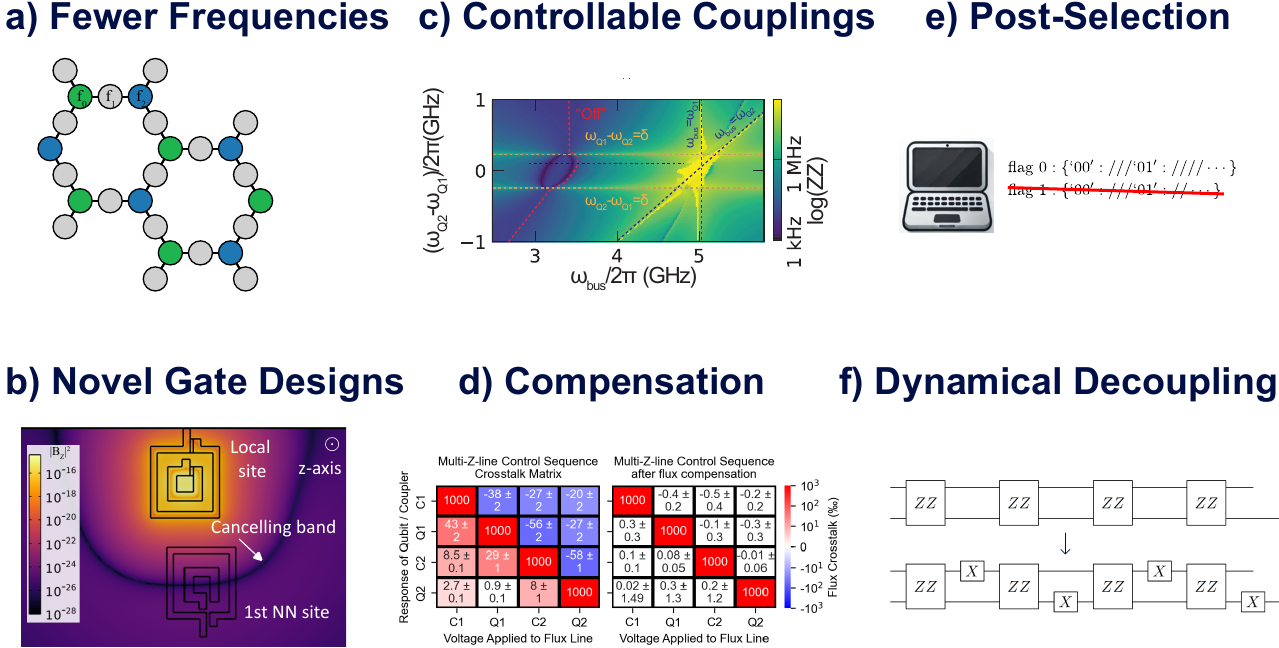}
    \caption{Crosstalk mitigation techniques. a) A heavy-hexagon lattice with connectivity of degree 3, requires only three unique frequencies so adjacent qubits do not share a frequency, reducing frequency collisions as described in Ref. \cite{Chamberland_20_topological}. b) Simulated variation of the $Z$ component of magnetic field strength for systems where control is performed with dual concentric loops \cite{Weng_24_crosstalk}. c) $ZZ$ interaction strength as a function of qubit-qubit detuning and bus frequency in a superconducting \gls{qpu} with tunable-couplers \cite{Stehlik_21_tunable}. d) Flux crosstalk between qubits and couplers before and after compensating for non-local flux when driving individual flux control lines in a superconducting \gls{qpu} \cite{Ma_25_characterizing}. e) Post-selection of data based on crosstalk detection circuit measurements \cite{Kang_25_time}. f) Staggered dynamical decoupling sequences targeted at suppressing always-on $ZZ$ interactions \cite{Niu_24_multi, Evert_24_syncopated}.}
    \label{fig:mitigation}
\end{figure*}

Crosstalk mitigation reduces the negative impact of crosstalk when running quantum circuits. In shared tenant environments, this can be desired to mitigate the risks associated with adversarial attacks, discussed in Section \ref{sec:Security}. In single-tenant environments, crosstalk mitigation can improve the execution of quantum circuits sampled in parallel or individual circuits when crosstalk contributes to a significant portion of overall noise. In the remainder of this section, we will review crosstalk mitigation as it appears in quantum computing discussions from as early as the \gls{qpu} design stage, influencing details such as qubit connectivity~\cite{Chamberland_20_topological} and tunability~\cite{Ding_20_Systematic}, to the circuit compilation stage for individual \gls{nisq} algorithms \cite{Seif_24_suppressing} and \gls{qec} circuits \cite{Cheng_24_crosstalk}. 

\subsection{Architectural}
Significant variety in crosstalk mitigation techniques is available at the architectural design stage. This includes changing how gates are applied, how measurements are performed and how qubits are spatially arranged or defined within the system Hilbert space. The next paragraphs detail device design approaches present in the literature which have consequences on the properties of crosstalk noise.

A consistent theme among platforms is to design systems to avoid components vulnerable to spillover crosstalk and frequency collisions. To this end, IBM initially developed larger devices with heavy-hexagonal connectivity~\cite{Chamberland_20_topological, Jurcevic_21_demonstration}, reducing distinct frequency requirements while minimising the likelihood of frequency collisions among adjacent qubits. Heavy-hexagonal connectivity in a smaller device is shown in Figure \ref{fig:mitigation}~a). IBM also developed a post-fabrication laser annealing technique to tune qubits to avoid frequency collisions and mitigate the severity of always-on $ZZ$ interactions~\cite{Zhang_20_high}.

Other variants of superconducting \glspl{qpu} have also been investigated in the context of crosstalk mitigation at the architectural level. For example, \gls{oqc} has investigated the logical performance of devices featuring dimon logical qubits~\cite{Wills_25_error}, achieving logical error rates which were an order of magnitude lower than physical error rates. Beyond this, arguments are made that dimon logical qubits are sensitive to the relative changes in physical mode frequencies, rather than to individual changes in mode frequency as seen in transmons. This allows in-built robustness to correlated frequency perturbations, similar to decoherence-free subspace techniques suggested for trapped-ion devices \cite{Kielpinski_02_architecture}. In parallel, \gls{oqc} are developing a scalable superconducting \gls{qpu} architecture using a transmon variant called coaxmons, which are optimised for crosstalk suppression and two-dimensional tiling. Results shown, before the inclusion of qubit coupling circuitry, confirm significant crosstalk suppression \cite{Spring_22_high}. 

Designs achieving frequency separation are also considered for \glspl{qpu} using identical particles such as atoms and ions. For example, when trapped ions are used as qubits in quantum communication nodes, high fidelity entanglement may be formed between trapped ions and emitted photons. The ions used to generate entangled ion-photon pairs, known as communication ions, can emit photons which unintentionally become absorbed with other ions used to store information, known as memory ions. Multiple ion species may be used to combat this phenomenon, such that the emitted photons from one species is not resonant with the other~\cite{Hughes_20_benchmarking}. Alternatively, this crosstalk mechanism can be mitigated using a single ion species if resonance is avoided by encoding communication and memory qubits in different hyperfine levels~\cite{huang_25_realization}.

Different mechanisms for applying quantum gates sometimes offer benefits in crosstalk reduction. In \gls{nv} \glspl{qpu}, instead of global microwave control and pursuing spectral resolution, Weng \etaldot investigated minimising microwave crosstalk by placing pairs of concentric square microwave sources at individual \glspl{nv}~\cite{Weng_24_crosstalk}, as shown in Figure \ref{fig:mitigation} b). Both microwave sources at each qubit were individually controllable, and able to be run with tunable relative amplitude and phase. Optimising these parameters resulted in the ability to minimise the crosstalk at the nearest-neighbour adjacent \gls{nv}. Another research study developed electric field control of qubits of NV centers by taking advantage of strain effects \cite{Wang_23_field}. Theoretical simulations suggest that this may offer opportunities to invert the negative effects of crosstalk due to the linear relationship between applied voltages and electric fields attained at different sites in the crystal lattice.

\subsection{Tuning}
Tuning device elements is usually the first approach to mitigate crosstalk after a device has been fabricated. This corresponds to techniques such as changing applied voltages, compensating for observed spillover effects and changing device component parameters to avoid undesired interactions. The next paragraphs detail how these can be applied towards the goal of mitigating crosstalk. A review of crosstalk-tolerant calibration of readout, which is performed classically and usually external to the \gls{qpu} is reserved to Section \ref{sec:Readout}.

Depending on the free parameters available in a superconducting \gls{qpu}, some always-on $ZZ$ interactions can be tuned away, such as by tuning couplers to a minimum interaction regime~\cite{Stehlik_21_tunable}, as shown in Figure \ref{fig:mitigation} c), or tuning pairs of couplers or other components so that contributions interfere destructively~\cite{Mundada_19_Suppression, Ku_20_suppression}. In the absence of tunable qubits or couplers, pulsed based methods can be used as a method to control the effective interactions between pairs of qubits~\cite{Wei_22_hamiltonian}. It should be noted that interactions between the higher energy levels of superconducting qubits are also relevant. It is generally challenging to tune away all interactions, with those among neglected higher energy levels and higher order effects likely remaining. Sung \etaldot~\cite{Sung_21_realization} optimise beyond a two-level framework for a system with a tunable coupler, reaching $T_1$-limited performance. 

Ai and Liu~\cite{Ai_25_scalable} investigated the use of graph neural networks to optimise frequency assignments in large superconducting systems. Optimised frequency assignments reduce frequency crowding. The method involves constructing two neural networks corresponding to an evaluator and a designer. Results show lower expected crosstalk error rates compared to the Snake optimiser, previously applied to Google processors~\cite{Klimov_20_Snake, Arute_19_quantum}.

In some cases, the control of tunable elements is itself prone to crosstalk. Ma \etaldot~\cite{Ma_25_characterizing} showcase this in their work developing the \gls{mzlc} protocol to mitigate magnetic flux crosstalk in a tunable transmon superconducting \gls{qpu}. The \gls{mzlc} protocol sweeps flux bias voltage and microwave control pulse frequency, identifying the microwave frequencies which result in peaks in the analogue-to-digital converter in-phase channel as a function of flux bias voltage. The procedure is repeated while varying the flux bias of an adjacent qubit or coupler in addition to original qubit or coupler under consideration. A linear relationship was observed between the pair of flux bias values which correspond to driving the system at the $0-1$ transition, suggesting flux bias was affecting multiple system elements. This linear relationship was used to construct a flux crosstalk matrix, as shown in Figure \ref{fig:mitigation} d), for which the inverse was then used to reconfigure flux pulses, yielding a reduction in flux control line crosstalk from $56.5\%$ to $0.13\%$. Trapped-ions experience a similar spillover phenomenon when implementing \gls{ms} gates. Ref.~\cite{Kashyap_25_Crosstalk} address this by finding linear combinations of single-mode couplings which results in zero couplings between target and neighbour ions.

Tuning can be a challenge for \glspl{qpu} composed of literally atomic components, due to some system parameters deriving from intrinsic properties rather than externally controllable parameters. In such cases, reducing crosstalk using physical phenomena may still be possible. Micromotion hiding is such a technique, used in Quantinuum systems~\cite{Gaebler_21_suppression}. Absorption of measurement light for non-measured ions is suppressed by applying a protocol that causes these ions to oscillate within their traps, causing a Doppler shift of absorption frequencies away from the measurement laser and measured ion fluorescence. Micromotion hiding is also compatible with the alternative technique of moving ions away to different gate zones, with results showing improved performance even if measured and idle ions are in different zones~\cite{Hothem_25_measuring}.

\subsection{Readout}\label{sec:Readout}
Crosstalk mitigation of readout operations can be distinguished from other mitigation problems by corresponding largely to classical data processing. Traditional approaches correspond to fitting a decision boundary based on calibration data. Solutions are ideally both accurate at identifying corrected measurement outcomes, and fast when subsequent calculations depend on mid-circuit measurement results, such as in \gls{qec} applications. Applications of machine learning techniques have seen significant interest in both these fronts \cite{Alexeev_25_Artificial}. Here we overview the literature which has investigated means of mitigating the measurement errors associated with readout crosstalk.

Readout in neutral atom \glspl{qpu} can be performed by applying resonant laser light to atoms and capturing the fluoresce response using a \gls{ccd} digital camera. The assignment of qubits in bright/dark states increases in difficulty as measurement times and inter-qubit spacings are reduced. Initial work investigated readout crosstalk mitigation utilised \glspl{cnn}~\cite{Phuttitarn_24_enhanced}, realising accuracy improvements but required increased computation cost and reduced interpretability. Subsequent work investigated the use of matched filters, which are simpler optimised weighted sums of intensity values \cite{kent_25_efficient}. Results found that optimal performance was obtained with filters resembling Gaussians with some deviations for pixels at the periphery of each qubit. When averages of other qubits were included as features, optimal weights were found to consistently be negative directly below and to the right of each qubit, indicating systematic imperfections causing fluorescence from neighbouring qubits to propagate in specific directions. Recently, Mude \etaldot \cite{Mude_25_Enabling} further develop on these techniques by investigating whether denoising can be used to allow lighter-weight machine models which offer reduced computation times during inference.

Machine learning techniques have also been developed to mitigate readout crosstalk of superconducting systems. Here crosstalk errors are introduced by frequency multiplexed readout, where the state of adjacent qubits can affect a particular readout result. Measurement data corresponds to two-dimensional signals in the In-phase (I) and Quadrature (Q) plane generated by analogue to digital converters.  Feed-forward neural networks have been shown to provide higher accuracy classification compared to traditional methods \cite{Lienhard_22_deep}. Performance advantages generally increased as more qubits required simultaneous readout classification, which were associated with correlated measurement errors due to crosstalk.

Latency and throughput of readout classification are of particular concern for superconducting \glspl{qpu} due to the significantly reduced gate time. Maurya \etaldot~\cite{Maurya_23_Scaling} developed a FPGA-compatible feed forward neural network for readout discrimination of superconducting qubits. Matched filters were used to reduce the size of the input layer. The trained ANN was able to mitigate the readout errors associated with these effects by learning the most appropriate label to assign given the trajectories observed in past data. Mude \etaldot~\cite{Mude_25_efficient} subsequently offered additional performance improvements, both in classification accuracy and latency.

\subsection{Compilation}
Compilation corresponds to the mapping of a high-level quantum circuit to low level instructions to be implemented on a \gls{qpu}. This includes making decisions about how each high-level instruction is decomposed, and where and when the resultant operations are performed. The location of computations is particularly relevant when significant inhomogeneity at different device locations, and can be a efficient means of noise mitigation at the compilation stage~\cite{Nation_23_Suppressing}. Moreover, the numerous different decompositions of circuit operations also offer a means to mitigate the effect of crosstalk noise during compilation. The following paragraphs focus on compilation techniques involving qubit placement and gate compilation, with echoing techniques discussed next in Section \ref{sec:Echoing}.

Optimised qubit allocation in the context of crosstalk noise was investigated by Harper \etaldot~\cite{Harper_24_crosstalk} using a reinforcement learning algorithm. A policy parametrised by a neural network was trained over 500,000 instances of random five-qubit circuits with up to 20 gates. Significant improvements in circuit fidelity were demonstrated when the learned policy was used for qubit allocation, with an ability to increase fidelity from 0.49 to 0.99, when compared with unoptimised qubit placement for a four-qubit circuit. Furthermore, generalisation to variation in noise conditions was also demonstrated by using partial re-training using data from the modified noise model.

Murali \etaldot~\cite{Murali_20_software} developed a compiler aimed at mitigating the impact of crosstalk due to simultaneous application of CNOT gates on IBM fixed frequency transmon devices. After first performing crosstalk characterisation, the scheduling (temporal placement) of gates is optimised to achieve a good balance between parallel execution, which is vulnerable to crosstalk, and serial execution, which increases circuit duration and hence decoherence. The compiler maps the optimisation problem to a form compatible with a satisfiability modulo solver. Results given show the compiler reduced the error rate of swap gates, QAOA circuits, and the hidden shift benchmark. Xi \etaldot \cite{Xie_21_Mitigating} extended this work by additional performing circuit gate reordering to maintain benefits of short circuit length.

The compilation technique by Seif \etaldot~\cite{Seif_24_suppressing} focused on addressing always-on and context dependent coherent errors in IBM superconducting devices. The mitigation of always-on $ZZ$ interactions was implemented by applying \gls{dd} pulses to qubits experiencing errors with distinct pulse schedules. Compensating operations are performed by absorbing single qubit $R_Z(\theta)$ errors into the Euler angle decomposition of subsequent single qubit gates and absorbing two-qubit $R_{ZZ}(\theta)$ errors into subsequent entangling operations such as Cartan decompositions, when possible. These methods were demonstrated to be able to remove oscillations which would otherwise occur during repeated application of test circuits, and improve circuit fidelity in Ising and Heisenberg model simulation circuits.

Detecting problematic circuits submitted to a shared quantum environment and an alternative means of reducing the impact of crosstalk during compilation. \cite{Deshpande_23_Design}, Deshpande \etaldot use a directed acyclic graph (DAG) circuit representation and pattern matching formulated as a subgraph isomorphism problem to create an algorithm for the malicious circuit detection they previously described. This algorithm was able to identify all malicious circuits the authors tested and only gave a false positive for a Bernstein–Vazirani circuit. This false positive included information that matched only one instance of an attacking circuit which is insufficient to cause algorithm failure, and so could potentially be avoided by thresholding the antivirus to only signal a malicious user when sufficiently many malicious circuits are detected. Resource estimates were given when this antivirus is run on a Intel(R) Xeon E-2286G CPU at 4GHz yielding times of approximately 250 seconds per tested circuit. While initial results are promising, the efficacy of this approach remains to be been tested in practice on a real device.

Finally, the crosstalk detection protocol of Kang \etaldot \cite{Kang_25_time}, which operates by the preparation of auxiliary crosstalk detection states, may also be used as a crosstalk mitigation technique when compiling quantum algorithms. As the technique can be performed in parallel to the execution of a useful quantum circuit, it also can be reduce the negative effect of crosstalk by post-selecting to only use data which did not trigger crosstalk detection events, as represented in Figure \ref{fig:mitigation} e). 

\subsection{Coherent Echoing}\label{sec:Echoing}
\glsreset{dd}
Many crosstalk mitigation techniques take advantage of coherence by using echoing or \gls{dd} pulses~\cite{Viola_99_dynamical} which cause parts of the noisy evolution to destructively interfere. A limitation of these techniques is that they require the availability of low-noise additional pulses, and require noise to not be completely incoherent and to not change quickly in time. The following paragraph summarises instances where such conditions are met when applied to crosstalk mitigation.

Standard \gls{dd} pulse sequences are usually designed to mitigate single-qubit coherent errors which are correlated in time, usually unable to suppress arbitrary multi-qubit coherent interactions. Tripathi \etaldot \cite{Tripathi_22_suppression} considered a special case where coherent interactions between particular qubits and spectator (non-utilised) qubits are instead targeted in particular. In this case standard $XY4$ \gls{dd} sequences, when applied to each spectator qubit, were able to successfully suppress always-on $ZZ$ crosstalk both when target qubits were kept idle and participating in cross-resonance gates on IBM hardware. Limitations to this approach were also observed, as in some cases single qubit coherent oscillations still remained, which were concluded to be due to calibration choices. However, the dependence of these these coherent effects on the spectator qubit states was successfully eliminated.

There has also been some effort to mitigate always-on $ZZ$ crosstalk for complete sets of qubits in quantum devices. Niu \etaldot~\cite{Niu_24_multi} investigated methods to extend this suppression to algorithmic qubits using staggered \gls{dd} sequences, as show in Figure \ref{fig:mitigation} f). Pairs of qubits were pulsed independently on disjoint time steps, successfully mitigating always-on crosstalk and driven (operational) crosstalk. Evert \etaldot~\cite{Evert_24_syncopated} address suppression of multi-qubit coherent interactions more generally, including a frequency doubling approach, targeting of specific interactions, and applications to systems with known crosstalk graphs. Methods were validated by achieving noise suppression of always-on $ZZ$ interactions in superconducting qubits. Zhou \etaldot~\cite{Zhou_23_quantum} also considered crosstalk suppression in the context of always-on $ZZ$ interactions in superconducting \glspl{qpu}. A perturbative expression for crosstalk errors valid in the regime of strong control and weak noise was identified. A condition was found identifying whether leading order contributions to noise were suppressed. A staggered $XY4$ \gls{dd} scheme was identified which met this condition, and achieved noise suppression applied to state-preparation and measurements and \gls{qns} circuits.

Coherent echoing of crosstalk noise has been used in the context of performing quantum gates. In cross-resonance superconducting \glspl{qpu}, this has been used to mitigate the pulse spillover on the target qubit in the form of undesired single qubit rotations during cross-resonance gates~\cite{Sheldon_16_procedure, Sundaresan_20_reducing}. Pulse spillover crosstalk has also been mitigated in trapped-ion during the application of two-qubit \gls{ms} gates \cite{Fang_22_crosstalk}. In this case, pulse spillover was mitigated by applying the two-qubit interaction pulses in terms of two separate pulses which cancel each other out, to first order. To allow the pulses to apply the \gls{ms} gate to the qubits of interest, the second pulse contains the addition of echoing $Z$ gates on participating qubits, which causes the two pulses to interfere constructively. The key enabling factor in this case was the ability to echo the pulse of the \gls{ms} gate with Pauli gates, although similar techniques may be possible on other architectures suffering pulse spillover. A similar method was investigated by Zhang \etaldot~\cite{Zhang_22_Hidden}, who instead took advantage of the self-adjoint property of many common quantum gates. Coherent errors of multi-qubit $Z$ rotation gates on an IonQ device were mitigated by choosing whether to apply either a gate regularly or an equivalent version composed of inverse operations. 

\section{\label{sec:Security} Crosstalk Security Vulnerabilities}

\begin{figure*}
    \centering
    \includegraphics[width=\textwidth]{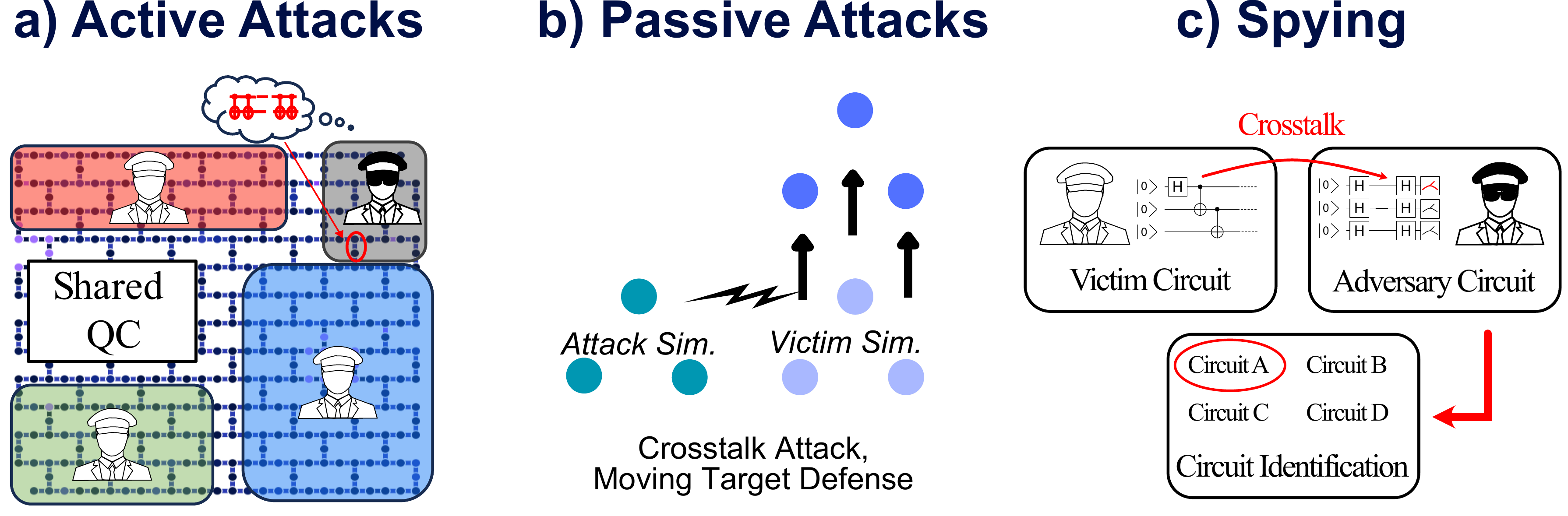}
    \caption{Security vulnerabilities enabled by crosstalk in \glspl{qpu}. a) An active attack, where a quantum computer tenant attacks a co-tenant using a sequence of repeated CNOT gates on superconducting hardware \cite{Harper_24_crosstalk}. b) A passive attack, where always-on interactions cause a victim simulation to fail due to an attack simulation which was placed too close \cite{sharma_25_evaluation}. c) Spying, where crosstalk detection using an adversarial quantum circuit facilitates victim circuit identification, based on descriptions given in Ref. \cite{choudhury_24_crosstalk}. }
    \label{fig:security}
\end{figure*}

Rapid \gls{qpu} device growth has spurred security concerns involving the use of quantum computers for computational tasks requiring privacy or reliability. Concern increased following a proposal for multi-tenant operation of cloud-based \glspl{qpu}~\cite{Das_19_case}. While the benefits of multi-tenant operation were argued under the restriction that providers controlled scheduling, the option of a utility computing model was also briefly suggested, which also gives scheduling access to users. As of yet, cloud-based \gls{qpu} providers still reserve control over circuit scheduling, with full multi-tenant access generally not offered.

Saki \etaldot \cite{Saki_21_survey} have previously given a review involving security concerns involving \glspl{qpu}, where they argue that the two main security concerns introduced by crosstalk are the potential for an adversary to cause a victim circuits to fail and to identify details of a victim circuit, as shown in Figure \ref{fig:security}. Interest regarding these topics continue in more recent research, which highlight novel `attack' vectors. The following sections review research which has investigated the sensitivity of quantum circuits to various crosstalk-mediated attack mechanisms.

\subsection{Pulse Spillover-Based Attacks}
Crosstalk noise can mediate pulse spillover attacks, where adversaries applying pulses to qubits they control can result in operations on victim qubits. These attacks are usually performed using circuits comprised of repeated single-qubit or two-qubit gates. The following paragraphs review the literature of crosstalk-mediated pulse-based attacks.

The most common pulse-based attack studied in the literature corresponds to cross-resonance pulse attacks on superconducting devices. Ash-Saki \etaldot~\cite{Ash-Saki_20_analysis} were the first to study the characterisation and modelling of this crosstalk on IBM fixed frequency transmon devices. Results revealed that CNOT crosstalk observed by spectator qubits showed a similar trend across different experimental runs. A model for noise during the application of CNOT gates was found using idle tomography~\cite{BlumeKohout_19_idle}. Experimental results showed that increased number of CNOT gates on qubits adjacent to a three qubit Grover's algorithm circuit caused measurement probabilities of solution states to become equal to those of non-solution states when approximately 30 attack CNOTs are applied. Results when attacking a quantum classifier circuit additionally showed that multiple attacks from different pairs of CNOTs were able to reduce the classifier accuracy further than each attack individually. Utilising buffer qubits to increase the minimum attack distance was found to reduce attack effectiveness. 

Bajpayee and Mukherjee~\cite{Bajpayee_24_Analysis} explored the sensitivity of IBM's 5 qubit devices to four different CNOT-based crosstalk attacks in small instances of Grover's algorithm, Simon's algorithm, and quantum teleportation. In all cases, attacks deteriorated device performance, though with some variation for each attack circuit which differs by different combinations of single qubit gates, identity gates, and delays applied to qubits adjacent to the victim circuit as it is executed in parallel.

Harper \etaldot~\cite{Harper_24_crosstalk} demonstrated spillover attack performance for a 3-qubit Grover's algorithm circuit on IBM superconducting hardware and simulations using noise models fit with idle tomography. Attack performance was investigated as a function of the distance from the victim circuit, defined using device connectivity. Results showed average victim fidelity reduces as separation from attack qubits decreases, though with significant variance in performance depending on precise attack qubit locations.  Evidence was also shown for the ability for CNOT crosstalk attacks to result in correlated dynamics within victim qubits. 

Mehra and Kalev \cite{Mehra_26_Towards} also explored the susceptibility of Grover's algorithm circuit to CNOT spillover crosstalk attacks on IBM hardware. Interestingly algorithm fidelity was found to depend on the number of attack CNOTs in a non-monotonic manner, as well as have dependence on the initialised state of the attack qubits. The defence mechanisms of using buffer qubits and dynamical decoupling were explored and found to give optimal performance when both are performed simultaneously.

Xu \etaldot~\cite{Xu_24_security} investigated whether malicious attacks at the pulse level can be applied and remain undetected depending on the verification capabilities available to a quantum device user. This highlights that, at the pulse level, errors consistent with crosstalk can resemble those of adversarial pulse attacks. Multiple attack modes are discussed, some taking advantage of crosstalk effects to perform non-local operations while the gate-level representation by the cloud provider may remain local. The circuits demonstrating vulnerabilities were quantum teleportation, Grover's search, and quantum neural networks. Tan \etaldot~\cite{Tan_25_Qubit} provided further demonstrations of the effectiveness on pulse spillover attacks using Qiskit Pulse on IBM and Rigetti superconducting devices. Four scenarios were tested varying an adversary's access to qubits and proximity to a victim circuit. In all cases, significant changes to the output distribution of the victim circuit were able to be achieved. These results showed that general quantum pulses, not simply two-qubit gates, can easily mediate pulse spillover attacks.

\subsection{Always-On Interaction Attacks}
Attacks utilising always-on-interactions apply minimal control pulses on attack qubits, instead taking advantage of the correlated dynamics inherent to the system Hamiltonian. As of the writing of this review, limited experimental results are available showing the vulnerability of different circuits to these types of attacks. The following paragraphs give examples of simulations of always-on-interaction attacks for superconducting and neutral atom hardware.

Shubha and Farheen~\cite{shubha_25_pulse} investigated the security implications of always-on interactions for superconducting qubits in a shared environment. Systems were simulated at the Hamiltonian level with the discrete time \textsc{mesolve} solver of QuTip~\cite{Li_22_pulse}. Case studies were considered where a victim had control of a single qubit, while an adversary had control of two adjacent qubits. Although only controlling single qubit interactions of adjacent qubits, it was found that nearest-neighbour always-on crosstalk, especially of the $YX$ and $ZX$ form, allowed significant modifications to the single qubit probability distribution of the victim. Attack strategies were considered where the attacker has pulse level control of adjacent qubits either exclusively before (attacker first) or exclusively after (victim first) the victim applies gates to prepare their intended single qubit state. For single qubit states prepared with continuous $Y$ rotations, the attacker first strategy was found to be significantly more effective. However, when attacking a single qubit XOR classification parametrised circuit, neither of the attack strategies were able to change output amplitudes beyond 1\%, revealing significant circuit dependence on the effectiveness of attacks mediated by always-on interactions.

Analog Hamiltonian simulator devices correspond to quantum devices which physically implement particular Hamiltonians, usually without support for universal quantum computation. Nevertheless, this computational model has seen success in areas such as optimisation, physics modelling, and molecular modelling~\cite{wurtz_24_industry}. Neutral atom arrays can be used as analog Hamiltonian simulators, where multi-qubit interactions occur due to van der Waals interactions causing the Rydberg blockade effect. Sharma \etaldot~\cite{sharma_25_evaluation} investigated the severity of crosstalk effects when analogue Hamiltonian simulations are performed in close proximity, using QuEra backends. Classical simulations modelling three qubits in an equilateral triangle formation were performed with separations in 1 $\mu \mathrm{m}$ increments from 4 $\mu \mathrm{m}$ to 7 $\mu \mathrm{m}$, showing a weak positive trend between circuit fidelity and distance between qubits of the two simulations. Interestingly, a minimum fidelity was observed at  5 $\mu \mathrm{m}$, causing the overall relationship to be non-monotonic. A \gls{mtd} strategy was suggested to defend against proximity-based adversarial attacks~\cite{Cho_20_toward}. This was found obtain fidelity of 0.995, compared to approximately 0.875 when no defence strategy is used.

\subsection{Crosstalk-Enabled Spying}
Spying corresponds to obtaining unauthorised information regarding the activities of a victim. Similar to the case when victim circuits are attacked, an adversary may apply no operations to their own qubits, but still look for signs of spillover from the victim's circuit. The following paragraphs give examples of research on the problem of distinguishing the circuits run in a shared environment based on the crosstalk noise experienced on spectator qubits, and identifying the previous states of qubits which have been reset. 

Choudhury~\etaldot~\cite{choudhury_24_crosstalk} demonstrated a method for crosstalk-mediated identification of quantum circuits in a multi-tenant model of cloud-based quantum computing . Experiments run on IBM superconducting \glspl{qpu} found that the probability of receiving non-zero measurement outcomes contained information correlated with the total numbers of CNOT gates in a circuit. Further arguments were given to suggest that sensitivity was in principle possible at the level of the total number of times each individual CNOT gate was applied, as well as the temporal locations of each CNOT gate. This motivated a method for identifying co-tenant circuits by utilising \glspl{gnn}, where the CNOT structure of circuit was given as input in the form of a graph. An accuracy of 85.7\% was achieved across a dataset of 336 circuits derived from MQT Bench~\cite{Quetschlich_23_mqtbench}. In terms of protection from such an attack, the authors suggest utilising a variety of qubit placements and circuit compilations to make the relationship between circuit and CNOT structure one-to-many. A major drawback of this spying strategy is that it requires a database of candidate circuits in order to identify what a victim is calculating.

Mi \etaldot~\cite{Mi_22_securing} investigated system vulnerabilities associated with reset gates on IBM 7-qubit superconducting \glspl{qpu}. Results found that qubits could be effectively reset on an individual basis, but are vulnerable to crosstalk-based identification of reset times. When victim qubits were prepared in single qubit states, it was found that non-secure reset operations, corresponding to a system-wide reset, resulted in qubits retaining correlations with the originally prepared states. These correlations were observed to remain after up to 31 repeated resets, performed by repeated preparation, measurement, and reset operations. An approach was developed to reduce such correlations by performing a random number of repeated resets, with idle padding added to avoid revealing the exact number based on circuit timing. This was found to reduce correlations to levels below those of system-wide resets, while maintaining a $300\times$ speedup compared to system-wide reset. However, it was also found that times when resets are performed on victim qubits can be identified using adjacent attack qubits (spectator qubits) initialised in $\ket{1}$ states. Specifically, Mi \etaldot argue that adversaries are able to identify the duration between the initialisation of a victim qubit and its last measurement operation, and the duration between the last measurement operation and the end of the victim's allocated share. 

\section{\label{sec:Outlook} Future Outlook}
As quantum platforms are still rapidly evolving, some topics relevant to crosstalk have yet to have widespread experimental investigation. These include topics involving new hardware developments, non-Markovianity, weaker crosstalk sources, and large-scale \gls{qec} systems. Here, we will review the literature available with respect to these topics, including speculative theoretical results.

\subsection{Non-Markovianity}
Standard descriptions of crosstalk usually begin by assuming Markovianity \cite{Sarovar_20_detecting}, where dynamics are expected to only depend on the current state of the system at any given time, without any memory of prior states. Non-Markovianity corresponds to dynamics that violates this. Examples include qubits interacting with some unknown bath that is not included in the computing model causing correlations in time and a dependence of future noise on which gates which have been applied in the past. As non-Markovianity has been shown to play a significant role in the noise profile of IBM quantum devices~\cite{White_20_demonstration}, it is natural to consider crosstalk topics in the presence of non-Markovian effects.

Proper treatment of non-Markovianity in crosstalk dynamics may significantly impact mitigation and characterisation protocols. Although general characterisation is likely to be costly, verifying and quantifying temporal violations of independence of operations may be computationally manageable. Generalisation of methods derived from studies of context-dependence may be helpful in this regard \cite{Rudinger_19_probing}.

Further understanding is needed on whether non-Markovianity introduces significant physical mechanisms which mediate crosstalk in systems interacting with an unknown bath. This is particularly relevant due to the inability to maintain a fixed state of such a bath, potentially causing long-lasting correlations in time. Studies of staggered \gls{dd} protocols may be valuable in this context \cite{Evert_24_syncopated}, where pulses can be engineered to only decouple qubits from an outside bath, retaining interactions between them. This may yield a protocol to determine the nature of the correlations between qubits, and eventually new mitigation strategies. Ideally this would result in rigorous classifications of instances of crosstalk as either bath-mediated, or genuine interaction between qubits.

Although it is well known that crosstalk forms a pathway for adversarial attacking through running a targeted circuit adjacent on a device, non-Markovianity may enable more complicated attack platforms. Current device performance may suggest that quantum non-Markovianity is unable to cause significant effects between subsequent experiments, as it would require unrealistic coherence. However, classical non-Markovianity may cause plausible and possibly significant effects. For both cases, further investigation is needed to understand the scope of attacks driven by non-Markovian effects.

\subsection{Crosstalk Sources}
Sparse study remains on exploring the physical phenomena which can cause crosstalk, or more generally correlated errors, in \gls{qpu} hardware. Further study can reveal novel noise sources to be kept in mind when performing resource estimates for large quantum algorithms, and may motivate sensing applications. As an example, consider the sensitivity of the spin of Rydberg atoms to gravitational fields. Neutral atom \glspl{qpu} usually spread Rydberg atoms across an array with controllable physical distance between qubits. This suggests a potential form of crosstalk mediated by changes in gravitational fields during a quantum algorithm. It would be interesting to estimate the conditions necessary to observe such effects. Moreover, this motivates the question of whether gravitational field gradients can induce (non-negligible) correlations within neutral atom \glspl{qpu}.  

With a thorough compilation of known crosstalk sources for different \gls{qpu} platforms, a checklist may then be constructed to verify their presence in new \glspl{qpu}. This could also be used to define hurdles to be demonstrated by \glspl{qpu} to ensure they are not vulnerable to known crosstalk mediated attack mechanisms. Even in single-tenant environments, where no security risks are of primary concern, this would provide valuable information for noise characterisation and mitigation. 

\subsection{Hardware Developments}
Developers of \glspl{qpu} regularly provide road-maps describing their plans for future quantum hardware and software. Here we will discuss road-map topics we found which may motivate the need for study from a crosstalk perspective.

Rigetti have tested floating tunable couplers for use in multi-chiplet quantum devices~\cite{Field_24_modular}. While present designs feature only two qubits, limiting what can be said about crosstalk but sufficient for \gls{rb}, researchers suspect that classical flux crosstalk is present when changing coupler flux bias. Net coupling values for multi-chiplet interactions were found to be comparable to those within a single chiplet.

IBM are planning on using long range quantum interconnects to support the implementation of \gls{qldpc} codes~\cite{Heya_25_randomized}. The characterisation of such devices may shed light onto how crosstalk can be expected to behave in superconducting systems with increased connectivity, such as those required for \gls{qldpc} codes \cite{Bravyi_24_High}.

All platforms are expecting to develop devices with significantly increased system sizes. This will almost certainly have implications for crosstalk noise strength and will intensify characterisation challenges. For example, IBM have stated that the shear quantity of pairs of qubits to consider limits what can be said using rigorous tests ~\cite{IBM_Eagle_Performance_2022}. An avenue for further work may be to develop algorithms to rigorously eliminate large sets of gates or circuit layers as crosstalk candidates.

\subsection{Quantum Error Correction}
The implementation of large-scale quantum algorithms is expected to require \gls{ftqc} supported by \gls{qec}. In this section we will review the literature on the role of coherent crosstalk in \gls{qec}, simulations of crosstalk error models for \gls{qec} systems, and discussions of crosstalk in experimental implementations of logical qubits. 

A major obstacle to investigating crosstalk in \gls{qec} systems is the difficulty of performing large-scale simulations with small coherent rotations. Such error models must nevertheless be studied as they are expected to appear experimentally and Pauli approximations underestimate the logical error rates that actually occur under coherent errors~\cite{Bravyi_18_Correcting}.  Fowler and Martinis~\cite{Fowler_14_quantifying} suggested that \gls{qec} syndrome measurements may reduce the impact of coherent spillover crosstalk. An argument was made that small correlated single-qubit rotations on multiple qubits would be transformed into independent Pauli errors after syndrome measurement, which would be mostly transformed into trivial identity operations due to the quantum Zeno effect~\cite{Misra_77_zeno}. However, in cases when correlated rotations are large, such as strong pulse spillover, \gls{qec} performance was expected to worsen. Further research is needed to understand when this transition occurs. In general, correlated errors caused by crosstalk are expected to reduce the effective code distance of \gls{qec} codes~\cite{Fowler_14_quantifying}. Additional understanding is needed to ensure that error suppression will persist at scales necessary for useful \gls{ftqc} algorithms~\cite{Chen_21_exponential}.

There are very few studies involving coherent crosstalk effects in \gls{qec}. Huang \etaldot~\cite{Huang_20_alibaba} investigated surface code sensitivity to $ZZ$ crosstalk in simulations based on superconducting \gls{qpu} systems. The crosstalk considered was always-on nearest-neighbour $ZZ$ interactions. When considering parameters based on devices available at that time, it was found that crosstalk leads to logical phase-flip error rates about 60\% higher than expected without crosstalk. For trapped-ion systems, phonon-mediated crosstalk during \gls{ms} gates in \gls{qec} circuits has been modelled using spin-dependent-force Hamiltonian calculations~\cite{Cheng_24_crosstalk, Liu_25_performance}. However, logical error rates were calculated using incoherent Pauli error models fit to samples of the coherent Hamiltonian calculations. The results for surface codes found that the two-qubit errors associated with crosstalk removed the possibility of a pseudothreshold in distance 3 codes. At larger code distances, the behaviour of logical qubit performance with increasing code distance depends significantly on the degree of parallelism used for multi-qubit gates and the speed at which gates are implemented. In the slow-gate regime, when phonons have time to propagate across multiple ions, an error threshold does not exist for large code distances, with logical error rates eventually beginning to increase with code distance. In the fast-gate regime, continuous error suppression behaviour returns, although correlated errors decaying polynomially with distance remain.

Interestingly, there has also been limited study of \gls{qec} with incoherent crosstalk. Debroy \etaldot~\cite{Debroy_20_Logical} investigated the implications of crosstalk noise in linear systems of trapped ions implementing \gls{qec} codes in the [[9,~1,~3]] compass code family. Crosstalk was modelled by applying correlated Pauli errors to qubits which are the nearest neighbours to those participating in \gls{ms} gates. Qubit mapping was optimised for each code by solving an equivalent Hamiltonian path problem, which was able to entirely eliminate distance-damaging crosstalk errors for the surface code. As expected from perfect solution to the Hamiltonian path problem, surface codes were found to perform best in regimes of high crosstalk. However, in regimes of low crosstalk, the Bacon-Shor and Shor codes were found to outperform surface codes. Zhou \etaldot~\cite{Zhou_25_surface} also investigated surface code performance with crosstalk using noise derived from IBM superconducting \gls{qpu} characterisation data. Incoherent multi-qubit $ZZ$ errors were placed after gates and idle periods. Results of memory circuits found that gate-based data-ancilla crosstalk caused the most significant increase in logical error rates, followed by gate-based data-data crosstalk. Stability circuit results found that gate-based data-ancilla crosstalk also degraded performance most significantly, however with less sensitivity to gate-based data-data crosstalk.

There are much more experimental \gls{qec} studies with direct comments to crosstalk. In a study by Google Quantum AI~\cite{Google_23_Supressing}, microwave crosstalk was considered a serious concern as it could contribute to leakage into higher energy states and was mitigated by applying compensating microwave tones. Simulations suggested that increases to $CZ$ crosstalk yielded the second greatest increase in logical error rate, behind only to the logical error rate increase resulting from increasing single qubit gate errors. IBM have characterised measurement crosstalk that occurs during the mid-circuit measurements of syndrome measurement circuits, finding that this can cause coherent $Z$ rotations between qubits that share readout resonators \cite{Chen_22_Calibrated}. Performing dynamical decoupling, centred in time on the measurement operation, on adjacent qubits was found to adequately address this form of crosstalk.

\section{\label{sec:Conclusion} Conclusion}
We reviewed crosstalk, its origin, and its character in different quantum computing architectures, as well as its security consequences. Reviewing crosstalk literature across different platforms revealed that focus on crosstalk remains limited in favour of other, currently more dominant, error mechanisms such as gate errors and independent decoherence. Each platform showed evidence of unique crosstalk mechanisms, such as control pulse spillover, always-on interactions, and frequency crowding. In terms of crosstalk characterisation and modelling, idle tomography remains the most popular~\cite{BlumeKohout_19_idle}, and is replaced with correlated Pauli error models when large-scale simulation is needed~\cite{Sarovar_20_detecting}. Crosstalk mitigation focused on destructively interfering coherent crosstalk sources, through techniques such as \gls{dd} or device tuning. Security vulnerabilities focus on the ability to cause an error on a victim qubit, cause a quantum algorithm to fail, or extract information regarding the operation of adjacent qubits in a multi-tenant environment. Defensive techniques focused on spacing qubits away from potential adversaries and preparing qubits in states sensitive to crosstalk detection~\cite{Harper_24_crosstalk, Kang_25_time}. Finally, we provided a set of potentially interesting topics for further exploration of crosstalk in the evolving quantum computing landscape. In particular, in \gls{qec}, coherent modelling of crosstalk effects are significantly lacking, highlighting the need for future attention \cite{Harper_26_Nonclifford}. Quantum computing is a rapidly developing technology, and in some cases also makes use of newly developed components and techniques. Hence, care must be taken that conclusions regarding use-case suitability from the perspective of crosstalk attacks continue to remain valid.\\

\begin{acknowledgments}
This work was funded by the Australian Army through Quantum Technology Challenge 2024. We thank Neil Dowling for discussions during the initial stages of this work.
\end{acknowledgments}

\appendix

\section{Theoretical Descriptions of Crosstalk}\label{sec:Models}
Experimentally, crosstalk noise arises due to insufficient isolation between qubits or insufficient individual addressability. While such effects can vary significantly between platforms, more general theoretical discussion of crosstalk is also present in the literature~\cite{Sarovar_20_detecting}. Theoretically, crosstalk noise sources correspond to violations of spatial locality or independence of operations. Violation of spatial locality occurs when qubits in a device can unintentionally interact with each other. Violation of independence of operations occurs when noise associated with a particular point in a quantum circuit depends on what is occurring at other locations. This is sometimes described as a `spillover' effect. 

Instances of crosstalk can be further separated into the categories of idle, operational, measurement, and preparation crosstalk~\cite{Rudinger_18_classifying}. Idle crosstalk is associated with qubits interacting during idle periods, where ideally they should undergo no evolution. Operational crosstalk refers to noise triggered as a byproduct of the deliberate application of another operation. Measurement and preparation crosstalk are special cases of operational crosstalk referring to additional noise on qubits when measurement or preparation operations are performed. 

Noise models which incorporate crosstalk often encounter difficulty scaling to large system sizes. The simplest approach to modelling crosstalk, which actually avoids this scaling issue, is to model crosstalk by increasing single-qubit or two-qubit error rates in general, resulting in essentially noisier independent noise models \cite{Google_25_observation}. This is motivated by the behaviour of simultaneous \gls{rb} \cite{Gambetta_12_characterization} and may suffice for accurate modelling of average results of circuits with multiple random contexts. In general, however, accurate modelling of crosstalk require direct modelling of effects which violate locality or independence of operations on an individual basis.

Violations of locality can be modelled by directly including noise channels corresponding to interaction Hamiltonians during idle periods \cite{Throckmorton_22_crosstalk}. Operational crosstalk can be modelled by incorporating non-local error channels when performing particular gates in a quantum circuit, associated with the characteristics of the physical mechanism behind the gate and the geometry of the \gls{qpu}. Models such as these can be fit to experimental data by performing general tomographic methods, such as idle tomography \cite{BlumeKohout_19_idle} or fitting non-Markovian noise models \cite{White_20_demonstration}.

In this article, the term ``crosstalk noise sources'' refers exclusively to the physical causes of correlated noise in a quantum device and the term ``crosstalk errors'' refers to the resultant impact on the logical information encoded in the qubits of a quantum device, consistent with prior work~\cite{Sarovar_20_detecting}. 

The consequences of crosstalk are error correlations which occur between qubits and quantum gates. These correlations may vary in spatial locality and ultimately lead to undesired correlated errors present in measurement statistics. However, it should be emphasised that the presence of correlated errors alone is not sufficient to conclude that crosstalk is present, as the correlated errors present may neither violate locality of operations nor individual addressability~\cite{Yoneda_23_noise, RojasArias_23_spatial}. It is only when the characteristics of these correlated errors depends on \glspl{qpu} operation that they correspond to crosstalk.

Finally, we will mention some characteristics of quantum device noise which can easily be confused with crosstalk. Within general context-dependence exists the possibility that the statistical behaviour of a quantum device varies as a function of time ~\cite{Rudinger_19_probing}. This is an example of drift, but not necessarily crosstalk. The high-weight errors which may be present after a native multi-qubit gate are also not considered crosstalk, as they can be understood as an error occurring only on the isolated subset of qubits involved in such operations. It should be noted that external causes of correlated errors are not included in the category of crosstalk, but are included in the category of correlated errors. An example of a correlated error source which is not associated with crosstalk is a cosmic ray impact \cite{Harrington_25_Synchronous}. Such an event would cause the occurrence of additional errors showing correlations in space and time, but such errors may not be associated with insufficient isolation between qubits or individual addressability.

\section{Open-Source Assets}\label{sec:Assets}
\subsection{Figure 1}
Quantum computing platforms.
\begin{enumerate}
    \renewcommand{\labelenumi}{\alph{enumi})}
    \item \href{http://creativecommons.org/licenses/by/4.0/}{CC BY 4.0}  \cite{Gambetta_17_building}.

    \item \href{http://creativecommons.org/licenses/by/4.0/}{CC BY 4.0} \cite{Chen_24_Benchmarking}.
    
    \item \href{http://creativecommons.org/licenses/by/4.0/}{CC BY 4.0}
\cite{Radnaev_25_universal}.

    \item \href{http://creativecommons.org/licenses/by/4.0/}{CC BY 4.0} \cite{Alexander_24_manufacturable}.
    
    \item \href{http://creativecommons.org/licenses/by/4.0/}{CC BY 4.0}
\cite{Tanttu_24_assessment}.
    
    \item \href{http://creativecommons.org/licenses/by/4.0/}{CC BY 4.0}
\cite{Oberg_25_bottomup}.
\end{enumerate}

\subsection{Figure 2}
Crosstalk phenomena.

\begin{enumerate}
    \renewcommand{\labelenumi}{\alph{enumi})}
    \item \href{http://creativecommons.org/licenses/by/4.0/}{CC BY 4.0} from the preprint at \href{https://arxiv.org/abs/2108.04530v2}{arXiv:2108.04530}, which was later published as Ref. \cite{Tripathi_22_suppression}.

    \item  \href{http://creativecommons.org/licenses/by/4.0/}{CC BY 4.0} \cite{ParradoRodriguez_21_crosstalk}. 

    \item \href{http://creativecommons.org/licenses/by/4.0/}{CC BY 4.0} \cite{Mude_25_Enabling}.

    \item  \href{http://creativecommons.org/licenses/by/4.0/}{CC BY 4.0} \cite{Fyrillas_25_Resource}.

    \item \href{http://creativecommons.org/licenses/by/4.0/}{CC BY 4.0} \cite{John_25_two}.

    \item \href{http://creativecommons.org/licenses/by/4.0/}{CC BY 4.0} \cite{Bradley_19_ten}.

\end{enumerate}

\subsection{Figure 3}
Crosstalk mitigation techniques.
\begin{enumerate}
    \renewcommand{\labelenumi}{\alph{enumi})}
    \item Original figure based on the description given in Ref. \cite{Chamberland_20_topological}. 

    \item \href{http://creativecommons.org/licenses/by/4.0/}{CC BY 4.0} \cite{Weng_24_crosstalk}.

    \item \href{http://creativecommons.org/licenses/by/4.0/}{CC BY 4.0} \cite{Stehlik_21_tunable}.

    \item \href{https://creativecommons.org/licenses/by-nc-sa/4.0/}{CC BY-NC-SA 4.0} \cite{Ma_25_characterizing}.

    \item Adapted with permission from authors of Ref. \cite{Kang_25_time}.

    \item Original figure based on descriptions given in Refs. \cite{Niu_24_multi, Evert_24_syncopated}.

\end{enumerate}

\subsection{Figure 4}
Crosstalk security vulnerabilities.
\begin{enumerate}
    \renewcommand{\labelenumi}{\alph{enumi})}
    \item Adapted from Ref. \cite{Harper_24_crosstalk} with permission from the authors.

    \item \href{http://creativecommons.org/licenses/by/4.0/}{CC BY 4.0} \cite{sharma_25_evaluation}.

    \item Original figure based on the description given in Ref. \cite{choudhury_24_crosstalk}. Assets from Ref. \cite{Harper_24_crosstalk} were used with permission from the authors. 

\end{enumerate}

\bibliographystyle{quantum_doi}

\bibliography{references}

\end{document}